\documentclass[11pt]{article}
\usepackage{amssymb}
\usepackage{colortbl}
\usepackage{amsfonts,amsmath, longtable}

\usepackage{comment}

        \def\theequation{\thesection.\arabic{equation}}

\newcommand{\tr}{{\rm tr}}

\newcommand{\mL}{{\mathcal L}}

\newcommand{\vf}{\varphi}
\newcommand{\al}{\alpha}
\newcommand{\be}{\beta}

\newcommand{\om}{\omega}
\newcommand{\vth}{\vartheta}

\newcommand{\bfq}{{\bf{q}}}
\newcommand{\bfp}{{\bf{p}}}

\newcommand{\Mat}{ {\rm Mat}(N,\mathbb C) }

\newcommand{\mC}{\mathbb C}
\newcommand{\mZ}{\mathbb Z}

\newtheorem{predl}{Proposition}

\def\beq{\begin{equation}}
\def\eq{\end{equation}}
\def\p{\partial}

\newtheorem{theor}{Theorem}

\newcommand{\mat}[4]{\left(\begin{array}{cc}{#1}&{#2}\\ \ \\{#3}&{#4}
\end{array}\right)}

\newcommand{\mats}[4]{\left(\begin{array}{cc}{#1}&{#2}\\ {#3}&{#4}
\end{array}\right)}

\def\res{\mathop{\hbox{Res}}\limits}

\begin{document}

\setcounter{page}{1}

\begin{center}

\

\vspace{0mm}

{\Large{\bf Elliptic Ruijsenaars-Toda and elliptic Toda chains:  }}


\vspace{3mm}

{\Large{\bf classical $r$-matrix structure and relation to XYZ chain }}




 \vspace{15mm}

 {\Large {D. Murinov}}
\qquad\quad\quad
 {\Large {A. Zotov}}

  \vspace{10mm}


%
 {\em Steklov Mathematical Institute of Russian
Academy of Sciences,\\ Gubkina str. 8, 119991, Moscow, Russia}


 {\small\rm {e-mails: murinov344@yandex.ru, zotov@mi-ras.ru}}

\end{center}

\vspace{0mm}

\begin{abstract}
We discuss the classical elliptic Toda chain introduced by Krichever
and the elliptic Ruijsenaars-Toda chain introduced by Adler, Shabat and Suris.
It is shown that these models can be obtained as particular cases
of the elliptic Ruijsenaars chain. We explain how the classical $r$-matrix structures
are derived for these chains. Also, as a by-product, we prove that the elliptic Ruijsenaars-Toda
chain is gauge equivalent to discrete Landau-Lifshitz model of XYZ type.
The elliptic Toda chain is also gauge equivalent to XYZ chain with special
values of the Casimir functions at each site.
\end{abstract}

%

{\small{ \tableofcontents }}

\bigskip\

\section{Introduction}\label{sec1}
\setcounter{equation}{0}
In \cite{KrToda} I. Krichever constructed an integrable chain called the elliptic Toda
chain\footnote{It was mentioned in \cite{Suris} that this model is contained in a classification
suggested by R. Yamilov in \cite{Yamilov}.}
It is described by the following Hamiltonian:
  \beq\label{w01}
  \begin{array}{c}
  \displaystyle{
H^{\hbox{\tiny{eToda}}}=-\frac12\sum\limits_{a=1}^n\Big(\log\frac{1}{\sinh^2(\bfp_a/2c)}+
\log\Big(\wp(\bfq_{a-1}-\bfq_a)-\wp(\bfq_{a-1}+\bfq_a)\Big)\Big)\,,
 }
 \end{array}
 \eq
where $\wp(x)$ is the Weierstrass elliptic function (\ref{a052}) and
and numeration of particles is modulo $n$: $\bfq_{n+1}=\bfq_1$,
$\bfq_0=\bfq_n$.
Equations of motion are as follows (here $c=-1/2$):
  \beq\label{w02}
  \begin{array}{r}
  \displaystyle{
\frac{{\ddot {\bf q}}_a}{{\dot {\bf q}}_a^2-1}=
E_1(\bfq_a-\bfq_{a-1})+E_1(\bfq_a+\bfq_{a-1})-E_1(2\bfq_a)+
}
\\ \ \\
  \displaystyle{
+E_1(\bfq_a-\bfq_{a+1})+E_1(\bfq_a+\bfq_{a+1})-E_1(2\bfq_a)\,,
  }
 \end{array}
 \eq
where
$E_1(x)=\vth'(x)/\vth(x)$ is the function (\ref{a07})\footnote{Authors
of \cite{Adler} use the Weierstrass $\zeta$-function instead of $E_1$.
 These two functions are related as given in
(\ref{a051}).}.
The elliptic Toda chain is a particular case of the elliptic
Ruijsenaars-Toda chain introduced by V. Adler and Yu. Suris in \cite{Adler}\footnote{Originally, this model
was introduced by V. Adler and A. Shabat in the rational form \cite{AdlerShabat}.}.
Equations of motion for this model were derived in the Newtonian form:
  \beq\label{w03}
  \begin{array}{c}
  \displaystyle{
\frac{2{\ddot {\bf q}}_a}{{\dot {\bf q}}_a^2-1}=
}
\\ \ \\
  \displaystyle{
={\dot {\bf q}}_{a+1}f({\bf q}_{a},{\bf q}_{a+1},\eta)
-{\dot {\bf q}}_{a-1}f({\bf q}_{a},{\bf q}_{a-1},\eta)+
g({\bf q}_{a},{\bf q}_{a+1},\eta)+g({\bf q}_{a},{\bf q}_{a-1},\eta)-4E_1(2{\bf q}_a)\,,
 }
 \end{array}
 \eq
where $\eta$ is a constant parameter and
  \beq\label{w04}
  \begin{array}{c}
  \displaystyle{
f(x,y,\eta)=E_1(x-y-\eta)+E_1(x+y+\eta)-E_1(x-y+\eta)-E_1(x+y-\eta)\,,
}
\\ \ \\
  \displaystyle{
g(x,y,\eta)=E_1(x-y-\eta)+E_1(x+y+\eta)+E_1(x-y+\eta)+E_1(x+y-\eta)\,.
 }
 \end{array}
 \eq
The equations for elliptic Toda model (\ref{w02}) follow from (\ref{w03}) in the case $\eta=0$ (then $f(x,y,0)=0$).
It is worth mentioning that there is a set of constants $\eta_k$ in \cite{Adler}. In this
paper we mainly study a ''homogeneous'' version when all $\eta_k$ are equal to $\eta$. The way
how to introduce a set of
arbitrary parameters $\eta_k$ is presented in Section \ref{sec44}.

{\bf Purpose of the paper.} Our first aim is to show that the elliptic Ruijsenaars-Toda chain (as well as the elliptic Toda chain) is described as a particular case of the elliptic ${\rm GL}_N$ Ruijsenaars chain on $n$ sites
constructed in \cite{ZZ}
(see also \cite{Nijhoff,Suris,AdlerLL,Adler} and references therein for constructions of discrete elliptic integrable systems).
Its Hamiltonian is defined as
  \beq\label{w14}
  \begin{array}{c}
  \displaystyle{
H^{\hbox{\tiny{eR}}}=c\sum\limits_{a=1}^n \log
\sum\limits_{j=1}^N
 \frac{\prod\limits_{l=1}^N\vth({ q}^a_j-{ q}^{a-1}_l-\eta) }
 {\vth(-\eta)\prod\limits_{l: l\neq j}^N\vth({ q}^{a}_j-{ q}^{a}_l) }\,e^{p^a_j/c}\,.
 }
 \end{array}
 \eq
where the momenta $p^a_j$ and coordinates $q^a_j$ are numerated by a pair of indices: $j=1,...,N$ and $a$ is the
number of site $a=1,...,n$. In the continuous non-relativistic limit this model turns into
1+1 integrable Calogero-Moser field theory \cite{Krich2,LOZ,Z24}. See also \cite{Z25} for 1+1 field theories
obtained by field generalizations of finite-dimensional elliptic integrable systems.

In the  $N=2$ case the Ruijsenaars chain (\ref{w14}) contains a pair of momenta $p^a_1$, $p^a_2$ and a pair
of coordinates $q^a_1$, $q^a_2$ at $a$-th site. Then we proceed to the same model but defined through
the
''center of mass'' coordinates at each site
\beq\label{w1400}
  \begin{array}{c}
  \displaystyle{
{\bf q}_a=\frac{q^a_1-q^a_2}{2}\,.
}
 \end{array}
 \eq
This possibility was argued in \cite{ZZ}. Then we have a single degree of freedom at each site.
It will be shown that the corresponding equations of motion exactly reproduce those for the
elliptic Ruijsenaars-Toda chain (\ref{w03}) (and therefore, those for the
elliptic Toda chain (\ref{w02}) when $\eta=0$).

Our second aim is to derive the classical $r$-matrix structures for the
elliptic Ruijsenaars-Toda and the elliptic Toda chains using the results of \cite{MuZ}, where the
classical $r$-matrix structure was derived for the
Ruijsenaars chain (\ref{w03}).
For this purpose, we
explain how the classical $r$-matrix structure for the Ruijsenaars chain changes when
the sums of coordinates are equal to zero at each site.
We also discuss the modified Lax representation, which is used for the elliptic Toda chain.
This modification
also changes $r$-matrix and provides a compact answer.

Finally, following \cite{ZZ}, we show that the
elliptic Ruijsenaars-Toda and the elliptic Toda chains are gauge equivalent to XYZ chain \cite{Skl2}.
A relation between the
elliptic Ruijsenaars-Toda and XYZ chain (discrete XYZ Landau-Lifshitz model) was also mentioned in \cite{Adler}.
We take advantage of the fact that the transition to the center of mass frame case
(when the sums of coordinates are equal to zero at each site) exactly corresponds
to factorization formulae for the Lax matrices. This allows to perform the gauge transformation and find
explicit change of variables.




\section{Elliptic ${\rm GL}_N$ Ruijsenaars chain}\label{sec2}
\setcounter{equation}{0}

\subsection{Brief description of the model}
 Let us recall construction of the periodic ${\rm GL}_N$ elliptic Ruijsenaars chain on $n$ sites \cite{ZZ}.
Its phase space $\mC^{2Nn}$ is parameterized by canonical coordinates
  \beq\label{w101}
  \begin{array}{c}
  \displaystyle{
 \{p^a_i,q_j^b\}=\delta^{ab}\delta_{ij}\,,\quad
 \{p^a_i,p_j^b\}=\{q^a_i,q_j^b\}=0\,,\quad i,j=1,...,N;\ a,b=1,...,n\,.
 }
 \end{array}
 \eq
The numeration of sites (the upper indices) is modulo $n$, that is
  \beq\label{w102}
  \begin{array}{c}
  \displaystyle{
 q^0_i= q^n_i\,,\ p^0_i= p^n_i\,,\quad
 q^{n+1}_i= q^1_i\,,\ p^{n+1}_i= p^1_i\,,\quad i=1,...,N\,.
 }
 \end{array}
 \eq
The monodromy matrix
  \beq\label{w103}
  \begin{array}{c}
  \displaystyle{
 T(z)=L^1(z)L^2(z)...L^n(z)\in\Mat
 }
 \end{array}
 \eq
is defined through the Lax matrices
 \beq\label{w104}
 \begin{array}{c}
  \displaystyle{
 L^a_{ij}(z)=
 \phi(z,{ q}^{a-1}_i-{ q}^{a}_j+\eta)
 \frac{\prod\limits_{l=1}^N\vth({ q}^a_j-{ q}^{a-1}_l-\eta) }
 {\vth(-\eta)\prod\limits_{l: l\neq j}^N\vth({ q}^{a}_j-{ q}^{a}_l) }\,e^{p^a_j/c}\,,
 \quad a=1,...,n;\quad i,j=1,...,N\,.
 }
 \end{array}
 \eq
The Hamiltonian $H$ comes from
\beq\label{w105}
 \begin{array}{c}
  \displaystyle{
 \exp(H/c)=\res\limits_{z=0}z^{n-1}\tr T(z)\,.
 }
 \end{array}
  \eq
  Each Lax matrix $L^a(z)$ has a single pole at $z=0$, that is
\beq\label{w1051}
 \begin{array}{c}
  \displaystyle{
\res\limits_{z=0}z^{n-1}\tr T(z)=\tr\Big(\res\limits_{z=0}L^1(z)\res\limits_{z=0}L^2(z)...\res\limits_{z=0}L^n(z)\Big)\,.
 }
 \end{array}
  \eq
   Due to (\ref{a04}) residue of each Lax matrix is a rank 1 matrix:
 \beq\label{w1052}
 \begin{array}{c}
  \displaystyle{
\res\limits_{z=0}L_{ij}^a(z)=\frac{\prod\limits_{l=1}^N\vth({ q}^a_j-{ q}^{a-1}_l-\eta) }
 {\vth(-\eta)\prod\limits_{l: l\neq j}^N\vth({ q}^{a}_j-{ q}^{a}_l) }\,e^{p^a_j/c}\,.
 }
 \end{array}
  \eq
  Therefore,
\beq\label{w106}
 \begin{array}{c}
  \displaystyle{
 H=c\sum\limits_{a=1}^n \log h_{a,a+1}\,,\qquad h_{n,n+1}=h_{n,1}\,,
 }
 \end{array}
  \eq
where
   \beq\label{w107}
 \begin{array}{c}
  \displaystyle{
 h_{a-1,a}=\sum\limits_{j=1}^N
 \frac{\prod\limits_{l=1}^N\vth({ q}^a_j-{ q}^{a-1}_l-\eta) }
 {\vth(-\eta)\prod\limits_{l: l\neq j}^N\vth({ q}^{a}_j-{ q}^{a}_l) }\,e^{p^a_j/c}\,.
 }
 \end{array}
  \eq
  The corresponding equations of motion (in the Newtonian form) take the form:
   \beq\label{w108}
 \begin{array}{c}
  \displaystyle{
  \frac{{\ddot q}^a_i }{ {\dot q}^a_i }=
   -\sum\limits_{l=1}^N {\dot { q}}_l^{a+1}E_1({ q}_i^a-{ q}_l^{a+1}+\eta)
  -\sum\limits_{l=1}^N {\dot { q}}_l^{a-1}E_1({ q}_i^a-{ q}_l^{a-1}-\eta)
  +2\sum\limits_{l\neq i}^N {\dot {q}}_l^{a}E_1({q}_i^a-{q}_l^{a})+
 }
 \\
   \displaystyle{
   +\sum\limits_{m,l=1}^N {\dot { q}}_m^{a}{\dot { q}}_l^{a+1}E_1({ q}_m^a-{ q}_l^{a+1}+\eta)
  -\sum\limits_{m,l=1}^N {\dot { q}}_l^{a}{\dot { q}}_m^{a-1}E_1({ q}_m^{a-1}-{ q}_l^a+\eta)\,.
   }
 \end{array}
  \eq
These equations are represented in
the form of (semi-discrete) Zakharov-Shabat equation:
\beq\label{w109}
 \begin{array}{c}
  \displaystyle{
  {{\dot L}^a}(z)=\{H,{ L}^a(z)\}={ L}^a(z){ M}^a(z)-{ M}^{a-1}(z){ L}^a(z)\,.
 }
 \end{array}
  \eq
Explicit expression for $M^a$ matrices can be found in \cite{ZZ}.

\subsection{Classical $r$-matrix structure}
The following statement was proved in \cite{MuZ}.
The Lax matrices $L^a(z)$ (\ref{w104}) satisfy the quadratic $r$-matrix structure
\begin{equation}\label{w130}
 \begin{array}{c}
   \displaystyle{
     c\{L_{1}^a(z),L_2^b(w)\}=
     \delta^{ab}\Big(L_1^b(z)L_2^b(w)r^b_{12}(z,w)-r_{12}^{b-1}(z,w)L_1^b(z)L_2^b(w)+
   }
     \\ \ \\
   \displaystyle{
     +L_1^b(z)s^{+,\,b}_{12}(z)L_2^b(w)-L_2^b(w)s_{12}^{-,\,b}(w)L_1^b(z)\Big)+
   }
     \\ \ \\
   \displaystyle{
     +\delta^{a,\,b-1}L_1^{b-1}(z)L_2^b(w)s_{12}^{-,\,b}(w)-
     \delta^{a,\,b+1}L_1^{b+1}(z)L_2^b(w)s_{12}^{+,\, b+1}(z)\,,
   }
 \end{array}
\end{equation}
where
\begin{equation}\label{w131}
 \begin{array}{c}
   \displaystyle{
    r_{12}^a(z,w)=
       }
     \\ \ \\
   \displaystyle{
    =\sum\limits_{i\neq j}^N\phi(z-w,q_i^a-q_j^a)E_{ij}\otimes E_{ji}
    +E_1(z-w)\sum\limits_{i=1}^N E_{ii}\otimes E_{ii}
    -\sum\limits_{i\neq j}^N E_1(q_i^a-q_j^a) E_{ii}\otimes E_{jj}\,,
    }
  \end{array}
\end{equation}
\begin{equation}\label{w132}
    s_{12}^{+,\,a}(z)=s_{12}^a(z)+u_{12}^{+,\,a}\,,
\qquad
    s_{12}^{-,\,a}(w)=s_{21}^a(w)-u_{12}^{-,\,a}\,,
\end{equation}
\begin{equation}\label{w134}
    u_{12}^{+,\,a}=\sum\limits_{i,j=1}^NE_1(q_j^{a-1}-q_i^a+\eta)E_{ii}\otimes E_{jj}\,,
\qquad
    u_{12}^{-,\,a}=-\sum\limits_{i,j=1}^NE_1(q_i^{a-1}-q_j^a+\eta)E_{ii}\otimes E_{jj}\,,
\end{equation}
and the matrices $s_{12}^a(z)$ are defined as
\begin{equation}\label{w136}
    L_1^a(z)s_{12}^a(z)=\sum\limits_{i,j=1}^N L_{ij}^a(z)
    \Big(E_1(z+q_i^{a-1}-q_j^a+\eta)-E_1(q_i^{a-1}-q_{j}^a+\eta)\Big)E_{ij}\otimes E_{ii}\,,
\end{equation}
\begin{equation}\label{w137}
    L_2^a(w)s_{21}^a(w)=\sum\limits_{i,j=1}^N L_{ij}^a(w)
    \Big(E_1(w+q_i^{a-1}-q_j^a+\eta)-E_1(q_i^{a-1}-q_{j}^a+\eta)\Big)E_{ii}\otimes E_{ij}\,.
\end{equation}
The delta-symbols $\delta^{a,\,b-1}$ and $\delta^{a,\,b+1}$ in (\ref{w130}) are defined modulo $n$
as in
(\ref{w102}).

Then the
monodromy matrix $T(z)$ (\ref{w103})
satisfies the following $r$-matrix structure:
\begin{equation}\label{w140}
 \begin{array}{c}
   \displaystyle{
     c\{T_1(z),T_2(w)\}=T_1(z)T_2(w)r_{12}^n(z,w)-\left(r_{12}^n(z,w)-
     \breve{s}_{12}^{+,1}(z,w)+\breve{s}_{12}^{-,1}(z,w)\right)T_1(z)T_2(w)
       }
     \\ \ \\
   \displaystyle{
   +T_1(z)\breve{s}_{12}^{-,1}(z,w)T_2(w)-T_2(w)\breve{s}_{12}^{+,1}(z,w)T_1(z)\,,
    }
  \end{array}
\end{equation}
where $r_{12}^n(z,w)$ is defined in (\ref{w131}) for $a=n$, and $\breve{s}_{12}^{\pm,1}(z,w)$
are obtained by conjugation
\begin{equation}\label{w141}
 \begin{array}{c}
   \displaystyle{
    \breve{s}_{12}^{+,1}(z,w)=L_1^1(z)s_{12}^{+,1}(z,w)\left(L_1^1(z)\right)^{-1}\,,
       }
     \\ \ \\
   \displaystyle{
    \breve{s}_{12}^{-,1}(z,w)=L_2^1(w)s_{12}^{-,1}(z,w)\left(L_2^1(w)\right)^{-1}
    }
  \end{array}
\end{equation}
from ${s}_{12}^{\pm,1}(z,w)$ defined in (\ref{w132}) for $a=1$.
It follows from the above statement that
\beq\label{w147}
  \begin{array}{c}
  \displaystyle{
\{\tr \left(T^k(z)\right),\tr \left(T^l(w)\right)\} =0\,.
}
 \end{array}
 \eq

\section{Coordinates in the center of mass frame}\label{sec3}
\setcounter{equation}{0}
Here we consider the same Ruijsenaars chain but with coordinates
\beq\label{w301}
  \begin{array}{c}
  \displaystyle{
{\bar q}^a_i=q_i^a-\frac{1}{N}\sum\limits_{k=1}^Nq_k^a\,,\quad a=1,...,n;\quad i=1,...,N
}
 \end{array}
 \eq
which means that
\beq\label{w302}
  \begin{array}{c}
  \displaystyle{
\sum\limits_{k=1}^N{\bar q}^a_k=0\quad \forall a=1,...,n\,.
}
 \end{array}
 \eq
This means that we deal with center of mass frames at each site.

\subsection{Lax matrices and equations of motion}
The Lax matrices have the same form as in (\ref{w104}) but with $q^a_k$ replaced with ${\bar q}^a_k$:
 \beq\label{w303}
 \begin{array}{c}
  \displaystyle{
 {\bar L}^a_{ij}(z)=
 \phi(z,{\bar q}^{a-1}_i-{\bar q}^{a}_j+\eta)
 \frac{\prod\limits_{l=1}^N\vth({\bar q}^a_j-{\bar q}^{a-1}_l-\eta) }
 {\vth(-\eta)\prod\limits_{l: l\neq j}^N\vth({\bar q}^{a}_j-{\bar q}^{a}_l) }\,e^{p^a_j/c}\,,
 \quad a=1,...,n;\quad i,j=1,...,N\,.
 }
 \end{array}
 \eq
Introduce notation
 \beq\label{w401}
 \begin{array}{c}
  \displaystyle{
 b^a_{j}=
 \frac{\prod\limits_{l=1}^N\vth({\bar q}^a_j-{\bar q}^{a-1}_l-\eta) }
 {\vth(-\eta)\prod\limits_{l: l\neq j}^N\vth({\bar q}^{a}_j-{\bar q}^{a}_l) }\,e^{p^a_j/c}\,,
 \quad a=1,...,n;\quad j=1,...,N\,.
 }
 \end{array}
 \eq
Then
 \beq\label{w402}
 \begin{array}{c}
  \displaystyle{
 {\bar L}^a_{ij}(z)=
 \phi(z,{\bar q}^{a-1}_i-{\bar q}^{a}_j+\eta) b^a_{j}\,,
 \quad a=1,...,n;\quad i,j=1,...,N\,.
 }
 \end{array}
 \eq
The Hamiltonian is defined similarly to (\ref{w106})-(\ref{w107}):
\beq\label{w403}
 \begin{array}{c}
  \displaystyle{
 {\bar H}=c\sum\limits_{a=1}^n \log {\bar h}_{a,a+1}\,,\qquad
  {\bar h}_{a-1,a}=\sum\limits_{j=1}^Nb_j^a=\sum\limits_{j=1}^N
 \frac{\prod\limits_{l=1}^N\vth({\bar q}^a_j-{\bar q}^{a-1}_l-\eta) }
 {\vth(-\eta)\prod\limits_{l: l\neq j}^N\vth({\bar q}^{a}_j-{\bar q}^{a}_l) }\,e^{p^a_j/c}\,.
 }
 \end{array}
  \eq
  Equations of motion have the form (see \cite{ZZ}):
   \beq\label{w4031}
 \begin{array}{c}
  \displaystyle{
 {\dot q}_i^a=\frac{\p {\bar H}}{\p p_i^a}=
 \frac{b_i^a}{{\bar h}_{a-1,a}}\,,
 }
 \end{array}
  \eq
$$
  \displaystyle{
\frac{1}{c}\, {\dot p}_i^a=-{\dot q}_i^a\sum\limits_{l=1}^N E_1({\bar q}^{a}_i-{\bar q}^{a-1}_l-\eta)
-\sum\limits_{l=1}^N {\dot q}_l^{a+1}E_1({\bar q}_i^a-{\bar q}_l^{a+1}+\eta)
+\sum\limits_{l:l\neq i}^N({\dot q}_i^a+{\dot q}_l^a)E_1({\bar q}_i^a-{\bar q}_l^a)+
 }
 $$
   \beq\label{w4032}
 \begin{array}{c}
   \displaystyle{
   +\frac1N\sum\limits_{l=1}^N {\dot q}_l^a \sum\limits_{m=1}^N
   E_1({\bar q}^{a}_l-{\bar q}^{a-1}_m-\eta)
   -\frac1N\sum\limits_{l=1}^N {\dot q}_l^{a+1}\sum\limits_{m=1}^N
   E_1({\bar q}_l^{a+1}-{\bar q}_m^a-\eta)\,.
   }
 \end{array}
  \eq
Notice that $\sum\limits_{i=1}^N{\dot q}_i^a=1$ for all $a=1,...,n$.
In the Newtonian form we have:
   \beq\label{w4033}
 \begin{array}{c}
  \displaystyle{
  \frac{{\ddot q}^a_i }{ {\dot q}^a_i }=
   -\sum\limits_{l=1}^N {\dot { q}}_l^{a+1}E_1({\bar q}_i^a-{\bar q}_l^{a+1}+\eta)
  -\sum\limits_{l=1}^N {\dot { q}}_l^{a-1}E_1({\bar q}_i^a-{\bar q}_l^{a-1}-\eta)
  +2\sum\limits_{l\neq i}^N {\dot {q}}_l^{a}E_1({q}_i^a-{q}_l^{a})+
 }
 \\
   \displaystyle{
   +\sum\limits_{m,l=1}^N {\dot { q}}_m^{a}{\dot { q}}_l^{a+1}E_1({\bar q}_m^a-{\bar q}_l^{a+1}+\eta)
  -\sum\limits_{m,l=1}^N {\dot { q}}_l^{a}{\dot { q}}_m^{a-1}E_1({\bar q}_m^{a-1}-{\bar q}_l^a+\eta)\,.
   }
 \end{array}
  \eq

\subsection{Classical $r$-matrix structure}
Let us formulate how the classical $r$-matrix structure (\ref{w130})-(\ref{w137})
changes when using the coordinates ${\bar q}^a_k$ instead of $q^a_k$.
\begin{theor}\label{th1}
 The Lax matrices $L^a(z)$ (\ref{w303}) satisfy the following quadratic $r$-matrix structure:
\begin{equation}\label{w320}
 \begin{array}{c}
   \displaystyle{
     c\{{\bar L}_{1}^a(z),{\bar L}_2^b(w)\}=
     \delta^{ab}\Big({\bar L}_1^b(z){\bar L}_2^b(w)r^b_{12}(z,w)-r_{12}^{b-1}(z,w){\bar L}_1^b(z){\bar L}_2^b(w)+
   }
     \\ \ \\
   \displaystyle{
     +{\bar L}_1^b(z)s^{+,\,b}_{12}(z){\bar L}_2^b(w)-{\bar L}_2^b(w)s_{12}^{-,\,b}(w){\bar L}_1^b(z)\Big)+
   }
     \\ \ \\
   \displaystyle{
     +\delta^{a,\,b-1}{\bar L}_1^{b-1}(z){\bar L}_2^b(w)s_{12}^{-,\,b}(w)-
     \delta^{a,\,b+1}{\bar L}_1^{b+1}(z){\bar L}_2^b(w)s_{12}^{+,\, b+1}(z)\,,
   }
 \end{array}
\end{equation}
where
\begin{equation}\label{w321}
 \begin{array}{c}
   \displaystyle{
    r_{12}^a(z,w)=
       }
     \\ \ \\
   \displaystyle{
    =\sum\limits_{i\neq j}^N\phi(z-w,{\bar q}_i^a-{\bar q}_j^a)E_{ij}\otimes E_{ji}
    +E_1(z-w)\sum\limits_{i=1}^N E_{ii}\otimes E_{ii}
    -\sum\limits_{i\neq j}^N E_1({\bar q}_i^a-{\bar q}_j^a) E_{ii}\otimes E_{jj}\,,
    }
  \end{array}
\end{equation}
\begin{equation}\label{w322}
    s_{12}^{+,\,a}(z)=s_{12}^a(z)+u_{12}^{+,\,a}\,,
\quad
    s_{12}^{-,\,a}(w)=s_{21}^a(w)-u_{12}^{-,\,a}\,,
\end{equation}
\begin{equation}\label{w324}
    u_{12}^{+,\,a}=\sum\limits_{i,j=1}^NE_1({\bar q}_j^{a-1}-{\bar q}_i^a+\eta)E_{ii}\otimes \Big(E_{jj}-\frac1N\,1_N\Big)\,,
\end{equation}
\begin{equation}\label{w325}
    u_{12}^{-,\,a}=-\sum\limits_{i,j=1}^NE_1({\bar q}_i^{a-1}-{\bar q}_j^a+\eta)\Big(E_{ii}-\frac1N\,1_N\Big)\otimes E_{jj}\,,
\end{equation}
and the matrices $s_{12}^a(z)$ are defined through
\begin{equation}\label{w326}
    {\bar L}_1^a(z)s_{12}^a(z)=\sum\limits_{i,j=1}^N {\bar L}_{ij}^a(z)
    \Big(E_1(z+{\bar q}_i^{a-1}-{\bar q}_j^a+\eta)-E_1({\bar q}_i^{a-1}-{\bar q}_{j}^a+\eta)\Big)
    E_{ij}\otimes \Big(E_{ii}-\frac1N\,1_N\Big)\,,
\end{equation}
\begin{equation}\label{w327}
    {\bar L}_2^a(w)s_{21}^a(w)=\sum\limits_{i,j=1}^N {\bar L}_{ij}^a(w)
    \Big(E_1(w+{\bar q}_i^{a-1}-{\bar q}_j^a+\eta)-E_1({\bar q}_i^{a-1}-{\bar q}_{j}^a+\eta)\Big)
    \Big(E_{ii}-\frac1N\,1_N\Big)\otimes E_{ij}\,.
\end{equation}
Again, the delta-symbols $\delta^{a,\,b-1}$ and $\delta^{a,\,b+1}$ in (\ref{w320}) are understood modulo $n$.
\end{theor}
The proof is tedious but straightforward. It is similar to the proof of the statement (\ref{w130})-(\ref{w137})
given in \cite{MuZ}.

The monodromy matrix ${\bar T}(z)$ (\ref{w103}) defined with the Lax matrices (\ref{w303}) satisfies
the same quadratic $r$-matrix relation (\ref{w140}), where $r_{12}^n(z,w)$ is defined as in (\ref{w321})
and similarly to (\ref{w141})
\begin{equation}\label{w341}
 \begin{array}{c}
   \displaystyle{
    \breve{s}_{12}^{+,1}(z,w)={\bar L}_1^1(z)s_{12}^{+,1}(z,w)\left({\bar L}_1^1(z)\right)^{-1}\,,
       }
     \\ \ \\
   \displaystyle{
    \breve{s}_{12}^{-,1}(z,w)={\bar L}_2^1(w)s_{12}^{-,1}(z,w)\left({\bar L}_2^1(w)\right)^{-1}
    }
  \end{array}
\end{equation}
with ${\bar L}^1(z)$ (\ref{w303}) and $s_{12}^{\pm,1}$ (\ref{w322}). Therefore,
\beq\label{w342}
  \begin{array}{c}
  \displaystyle{
\{\tr \left({\bar T}^k(z)\right),\tr \left({\bar T}^l(w)\right)\} =0\,,\qquad
{\bar T}(z)={\bar L}^1(z)...{\bar L}^n(z)\,.
}
 \end{array}
 \eq

\subsection{Factorization and gauge equivalence with the higher rank Landau-Lifshitz XYZ chain}\label{sec33}
The Ruijsenaars chain was derived in \cite{ZZ} in two different ways. One possibility is to consider
elliptic solutions of 2d Toda lattice. This provides the ${\rm GL}_N$ model (\ref{w104})-(\ref{w108}).
Another way is to perform certain gauge transformation with the higher rank Landau-Lifshitz XYZ chain.
This construction assumes exactly the center of mass conditions (\ref{w302}). In this way one
comes to (\ref{w303})-(\ref{w4033}).

Introduce the following matrix \cite{Baxter2}:
  \beq\label{w351}
  \begin{array}{l}
  \displaystyle{
 g(z,q^a)=\Xi(z,q^a)\left(d^{a}\right)^{-1}\in\Mat
 }
 \end{array}
 \eq
 with
 \beq\label{w352}
 \begin{array}{c}
  \displaystyle{
\Xi_{ij}(z,q^a)=
 \vth\left[  \begin{array}{c}
 \frac12-\frac{i}{N} \\ \frac N2
 \end{array} \right] \left(z-N{\bar q}^a_j\left.\right|N\tau\right)\,,\quad i,j=1,...,N
 }
 \end{array}
 \eq
 and
 \beq\label{w353}
 \begin{array}{c}
  \displaystyle{
d^a_{ij}(z,q^a)=\delta_{ij}
 {\prod\limits_{k:k\neq j}^N\vth({\bar q}^a_j-{\bar q}^a_k)}\,,
 }
 \end{array}
 \eq
 where the definition of theta function (\ref{a02}) is used.
This is the intertwining matrix entering the IRF-Vertex
 correspondence in quantum statistical models \cite{Baxter2}.
 Its geometrical meaning and application to finite-dimensional
 integrable systems in classical mechanics can be found in \cite{LOZ,VZ}.

We are going to use the factorization property proved in \cite{Has}:
 \beq\label{w354}
 \begin{array}{c}
  \displaystyle{
 \Big( -\vth'(0)\, g^{-1}(z,{\bar q}^{a-1})g(z+N\eta,{\bar q}^a)\Big)_{ij}=
 \phi(z,{\bar q}^{a-1}_i-{\bar q}^{a}_j+\eta)
 \frac{\prod\limits_{l=1}^N\vth({\bar q}^a_j-{\bar q}^{a-1}_l-\eta) }
 {\prod\limits_{l: l\neq j}^N\vth({\bar q}^{a}_j-{\bar q}^{a}_l) }\,,
 }
 \end{array}
 \eq
that is we have the following factorized form for the Lax matrix ${\bar L}^a(z)$ (\ref{w303}):
 \beq\label{w355}
 \begin{array}{c}
  \displaystyle{
{\bar L}^a(z)=\frac{\vth'(0)}{\vth(\eta)}\,g^{-1}(z,{\bar q}^{a-1})g(z+N\eta,{\bar q}^a)\,e^{p^a/c}\,,
 }
 \end{array}
 \eq
where $e^{p^a/c}$ is a diagonal matrix with entries $e^{p^a_j/c}$.

The gauge transformation yields
 \beq\label{w356}
 \begin{array}{c}
  \displaystyle{
{\bar L}^a(z)\rightarrow g(z,{\bar q}^{a-1}){\bar L}^a(z)g^{-1}(z,{\bar q}^{a})=
\frac{\vth'(0)}{\vth(\eta)}\,
g(z+N\eta,{\bar q}^a)\,e^{p^a/c}\,g^{-1}(z,{\bar q}^{a})\stackrel{def}{=}{\mL}^a(z)
 }
 \end{array}
 \eq
Namely, consider the monodromy matrix ${\bar T}(z)$ (\ref{w342}) with ${\bar L}$-matrices in the factorized form:
 \beq\label{w3561}
 \begin{array}{c}
  \displaystyle{
{\bar T}(z)=\Big(\frac{\vth'(0)}{\vth(\eta)}\Big)^n
g^{-1}(z,{\bar q}^{n})g(z\!+\!N\eta,{\bar q}^1)\,e^{p^1/c}\cdot
g^{-1}(z,{\bar q}^{1})g(z\!+\!N\eta,{\bar q}^2)\,e^{p^2/c}...
}
\\ \ \\
  \displaystyle{
...g^{-1}(z,{\bar q}^{n-2})g(z\!+\!N\eta,{\bar q}^{n-1})\,e^{p^{n-1}/c}\cdot
g^{-1}(z,{\bar q}^{n-1})g(z\!+\!N\eta,{\bar q}^n)\,e^{p^n/c}\,.
 }
 \end{array}
 \eq
The gauge transformation (\ref{w356}) means
 \beq\label{w3562}
 \begin{array}{c}
  \displaystyle{
{\bar T}(z)\rightarrow {\mathcal T}(z)=g(z,{\bar q}^{n}){\bar T}(z)g^{-1}(z,{\bar q}^{n})\,,
 }
 \end{array}
 \eq
that is
 \beq\label{w3563}
 \begin{array}{c}
  \displaystyle{
{\mathcal T}(z)=\Big(\frac{\vth'(0)}{\vth(\eta)}\Big)^n
g(z\!+\!N\eta,{\bar q}^1)\,e^{p^1/c}
g^{-1}(z,{\bar q}^{1})\cdot g(z\!+\!N\eta,{\bar q}^2)\,e^{p^2/c}g^{-1}(z,{\bar q}^2)...
}
\\ \ \\
  \displaystyle{
...g(z\!+\!N\eta,{\bar q}^{n-1})\,e^{p^{n-1}/c}g^{-1}(z,{\bar q}^{n-1})
\cdot g(z\!+\!N\eta,{\bar q}^n)\,e^{p^n/c}g^{-1}(z,{\bar q}^{n})=
 }
 \\ \\ \\
   \displaystyle{
=\mL^1(z)\mL^2(z)...\mL^{n-1}(z)\mL^n(z)\,.
 }
 \end{array}
 \eq

The gauge transformed Lax matrix ${\mL}^a(z)$ (\ref{w356}) can be written in the form
of the (Sklyanin's type) Lax matrix for the higher rank Landau-Lifshitz model:
 \beq\label{w357}
 \begin{array}{c}
  \displaystyle{
  \mL^a(z)=\sum\limits_{\gamma\in\,\mZ_{ N}\times\mZ_{ N}} T_\gamma { S}^a_\gamma\exp(\frac{2\pi\imath \gamma_2 z}{N})\phi(z,\om_\gamma+\eta)\,,
  \quad \om_\gamma=\frac{\gamma_1+\gamma_2\tau}{N}\,,
 }
 \end{array}
 \eq
 where $T_\gamma$ is a special matrix basis in $\Mat$. For $\gamma=(\gamma_1,\gamma_2)\in\mZ_{ N}\times\mZ_{ N}$
 \beq\label{w358}
 \begin{array}{c}
  \displaystyle{
 T_\gamma=T_{\gamma_1 \gamma_2}=\exp\left(\frac{\pi\imath}{{ N}}\,\gamma_1
 \gamma_2\right)Q_1^{\gamma_1}Q_2^{\gamma_2}\,,\quad
  (Q_1)_{kl}=\delta_{kl}\exp(\frac{2\pi
 \imath}{{ N}}k)\,,\ \ \
 (Q_2)_{kl}=\delta_{k-l+1=0\,{\hbox{\tiny{mod}}}\,
 { N}}\,,
 }
\\ \ \\
  \displaystyle{
 T_\al T_\be=\kappa_{\al,\be}T_{\al+\be}\,,\quad \kappa_{\al,\be}=\exp\Big(\frac{\pi\imath}{N}(\be_1\al_2-\be_2\al_1)\Big)\,.
 }
 \end{array}
 \eq
 The coefficients $S^a_\gamma=S^a_\gamma(p^a,q^a,\eta)$
 are explicitly expressed through the variables $p^a_i$, $q^a_j$:
  \beq\label{w359}
  \begin{array}{c}
  \displaystyle{
 S^a_\gamma(p^a,q^a,\eta)=\frac{(-1)^{\gamma_1+\gamma_2}}{N}\,e^{\pi\imath \gamma_2\om_\gamma}
 \frac{\vth(\eta+\om_\gamma)}{\vth(\eta)}\sum\limits_{m=1}^N e^{p^a_m/c} e^{2\pi\imath \gamma_2(\eta-{\bar q}^a_m)}
 \prod\limits_{l:\,l\neq m}^N
 \frac{\vth({\bar q}^a_m\!-\!{\bar q}^a_l\!-\!\eta\!-\!\om_\gamma)}{\vth({\bar q}^a_m-{\bar q}^a_l)}\,.
 }
 \end{array}
 \eq
These are the generators of the classical ${\rm GL}_N$ Sklyanin algebra
\beq\label{w360}
 \begin{array}{c}
   \displaystyle{
 \{S^a_\al,S^a_\be\}=
 }
 \end{array}
  \eq
  $$
  \displaystyle{
 =\frac{1}{Nc}\sum\limits_{\xi\in\mZ_N^{\times 2},\,\xi\neq 0,\al-\be} \kappa_{\al-\be,\xi}S^a_{\al-\xi}S^a_{\be+\xi}
 \Big( E_1(\om_\xi)-E_1(\om_{\al-\be-\xi})+E_1(\om_{\al-\xi}+\eta)-E_1(\om_{\be+\xi}+\eta) \Big)
 }
$$
generated by the standard quadratic $r$-matrix structure
 \beq\label{w361}
 \begin{array}{c}
  \displaystyle{
 \{\mL_1^{a}(z),\mL_2^{a}(w)\}=\frac{1}{c}\,[\mL_1^{a}(z)\mL_2^{a}(w),r_{12}(z-w)]
 }
 \end{array}
 \eq
 with the classical elliptic Belavin-Drinfeld $r$-matrix \cite{BD}
 \beq\label{w362}
 \begin{array}{c}
  \displaystyle{
 r_{12}(z)=\frac{1}{N}\,1_N\otimes 1_N\, E_1(z)
 +\frac{1}{N}\sum\limits_{\gamma\in\mZ_N^{\times 2},\gamma\neq 0} T_\gamma\otimes T_{-\gamma}
 \exp(\frac{2\pi\imath \gamma_2 z}{N})\phi(z,\om_\gamma)
 \in{\rm Mat}(N,\mC)^{\otimes 2}\,.
 }
 \end{array}
  \eq
  Here $1_N$ is $N\times N$ identity matrix.

\section{Elliptic Ruijsenaars-Toda chain}\label{sec4}
\setcounter{equation}{0}

Here we study the Ruijsenaars chain (\ref{w303})-(\ref{w4033}) in the center of mass frame at each site
for the case $N=2$. Then all the Lax matrices and the Hamiltonian depend on coordinates
\beq\label{w471}
  \begin{array}{c}
  \displaystyle{
{\bar q}^a_1=-{\bar q}^a_2=\frac{q^a_1-q^a_2}{2}\,,\quad a=1,...,n\,,
}
 \end{array}
 \eq
that is, this model contains $n$ degrees of freedom. Since the Hamiltonian (\ref{w403}) depends on ${\bar q}^a_j$
we also have a set of conservation laws
\beq\label{w472}
  \begin{array}{c}
  \displaystyle{
p^a_1+p^a_2={\rm const}^a\,,\quad a=1,...,n\,,
}
 \end{array}
 \eq
and we choose ${\rm const}^a=0$ for all $a=1,...,n$. Introduce the following canonical variables
\beq\label{w473}
  \begin{array}{c}
  \displaystyle{
\bfq_a={\bar q}^a_1=\frac{q^a_1-q^a_2}{2}=-{\bar q}^a_2\,,
}
\\ \ \\
\bfp_a=2{p^a_1}=-{2}{p^a_2}\,,
 \end{array}
 \eq
i.e.
\beq\label{w474}
  \begin{array}{c}
  \displaystyle{
\{\bfp_a,\bfq_b\}=\delta_{ab}\,,\qquad
\{\bfp_a,\bfp_b\}=\{\bfq_a,\bfq_b\}=0\,,\quad a,b=1,...,n\,.
}
 \end{array}
 \eq

\subsection{Hamiltonian and equations of motion}
Plugging the definitions of new variables (\ref{w473}) into
the Lax matrices (\ref{w303}) and the functions (\ref{w401}) we obtain
\beq\label{w475}
  \begin{array}{c}
  \displaystyle{
{\bar L}^a(z)=\mat{\phi(z,\bfq_{a-1}-\bfq_a+\eta)b^a_1}{\phi(z,\bfq_{a-1}+\bfq_a+\eta)b^a_2}
{\phi(z,-\bfq_{a-1}-\bfq_a+\eta)b^a_1}{\phi(z,-\bfq_{a-1}+\bfq_a+\eta)b^a_2}\,,
}
 \end{array}
 \eq
where
\beq\label{w476}
  \begin{array}{c}
  \displaystyle{
b^a_1=\frac{\vth(\bfq_a-\bfq_{a-1}-\eta)\vth(\bfq_a+\bfq_{a-1}-\eta)}{\vth(-\eta)\vth(2\bfq_a)}
\exp\Big(\frac{\bfp_a}{2c}\Big)\,,
}
\\ \ \\
  \displaystyle{
b^a_2=-\frac{\vth(\bfq_a-\bfq_{a-1}+\eta)\vth(\bfq_a+\bfq_{a-1}+\eta)}{\vth(-\eta)\vth(2\bfq_a)}
\exp\Big(-\frac{\bfp_a}{2c}\Big)\,.
}
 \end{array}
 \eq
%
 Below we show that the equations of motion arising from the above Lax represnetaion coincide with equations of motion for the elliptic Ruijsenaars-Toda model
 (\ref{w03}) introduced in
\cite{Adler} for the case of pairwise equal $\eta_a=\eta_b=\eta$. The case of different $\eta_a$ is considered in the end of the Section.

 The Hamiltonian is as follows\footnote{Notice that we do not put the coefficient $c$ behind the sum in (\ref{w477})
 in contrast to (\ref{w403}). }
\beq\label{w477}
  \begin{array}{c}
  \displaystyle{
H^{\hbox{\tiny{eRT}}}=\sum\limits_{a=1}^n \log h^{\hbox{\tiny{eRT}}}_{a-1,a}\,,
\qquad h^{\hbox{\tiny{eRT}}}_{a-1,a}=b_1^a+b_2^a\,,
}
 \end{array}
 \eq
where the index $a$ is defined modulo $n$, that is $b_i^0=b_i^n$, $b_i^{n+1}=b_i^1$. The Hamiltonian
equations of motion take the following form:
\beq\label{w478}
  \begin{array}{c}
  \displaystyle{
{\dot \bfq}_a=\p_{\bfp_a}H^{\hbox{\tiny{eRT}}}=\frac{1}{2c}\frac{b_1^a-b_2^a}{b_1^a+b_2^a}\,,
}
 \end{array}
 \eq
\beq\label{w479}
  \begin{array}{c}
  \displaystyle{
{\dot \bfp}_a=-\p_{\bfq_a}H^{\hbox{\tiny{eRT}}}=
}
\\ \ \\
  \displaystyle{
-\frac{b_1^a}{b_1^a+b_2^a}\Big(E_1(\bfq_a-\bfq_{a-1}-\eta)+E_1(\bfq_a+\bfq_{a-1}-\eta)-2E_1(2\bfq_a)\Big)-
}
\\ \ \\
  \displaystyle{
-\frac{b_2^a}{b_1^a+b_2^a}\Big(E_1(\bfq_a-\bfq_{a-1}+\eta)+E_1(\bfq_a+\bfq_{a-1}+\eta)-2E_1(2\bfq_a)\Big)-
}
\\ \ \\
  \displaystyle{
-\frac{b_1^{a+1}}{b_1^{a+1}+b_2^{a+1}}\Big(E_1(\bfq_a-\bfq_{a+1}+\eta)+E_1(\bfq_a+\bfq_{a+1}-\eta)\Big)-
}
\\ \ \\
  \displaystyle{
-\frac{b_2^{a+1}}{b_1^{a+1}+b_2^{a+1}}\Big(E_1(\bfq_a-\bfq_{a+1}-\eta)+E_1(\bfq_a+\bfq_{a+1}+\eta)\Big)\,.
}
 \end{array}
 \eq
Let us compute the Newtonian form for the equations of motion.
By differentiating (\ref{w478}) with respect to time variable we get
\beq\label{w480}
  \begin{array}{c}
  \displaystyle{
{\ddot \bfq}_a=\frac{{\dot b}_1^a-{\dot b}_2^a}{b_1^a-b_2^a}\,{\dot \bfq}_a-\frac{{\dot b}_1^a+{\dot b}_2^a}{b_1^a+b_2^a}\,{\dot \bfq}_a=2{\dot \bfq}_a\,
\frac{ \frac{{\dot b}_1^a}{b_1^a}-\frac{{\dot b}_2^a}{b_2^a} }{ \frac{{b}_1^a}{b_2^a}-\frac{{b}_2^a}{b_1^a} }\,.
}
 \end{array}
 \eq
From (\ref{w478}) we also conclude that
\beq\label{w481}
  \begin{array}{c}
  \displaystyle{
\frac{{b}_1^a}{b_2^a}=\frac{1+2c{\dot {\bf q}}_a}{1-2c{\dot {\bf q}}_a}\,.
}
 \end{array}
 \eq
Therefore,
\beq\label{w482}
  \begin{array}{c}
  \displaystyle{
\frac{{b}_1^a}{b_2^a}-\frac{{b}_2^a}{b_1^a} =\frac{8c{\dot {\bf q}}_a}{1-4c^2{\dot {\bf q}}_a^2}\,.
}
 \end{array}
 \eq
Plugging it into (\ref{w480}) we get
\beq\label{w483}
  \begin{array}{c}
  \displaystyle{
\frac{4c{\ddot {\bf q}}_a}{1-4c^2{\dot {\bf q}}_a^2}=\frac{{\dot b}_1^a}{b_1^a}-\frac{{\dot b}_2^a}{b_2^a}\,.
}
 \end{array}
 \eq
The r.h.s. of (\ref{w483}) is computed directly using the definitions of $b_1^a$, $b_2^a$ (\ref{w476}):
\beq\label{w484}
  \begin{array}{c}
  \displaystyle{
\frac{{\dot b}_1^a}{b_1^a}-\frac{{\dot b}_2^a}{b_2^a}=
\frac{1}{c}\,{\dot {\bf p}}_a
+( {\dot {\bf q}}_a-{\dot {\bf q}}_{a-1} )\Big( E_1(\bfq_a-\bfq_{a-1}-\eta)-E_1(\bfq_a-\bfq_{a-1}+\eta) \Big)+
}
\\ \ \\
  \displaystyle{
+( {\dot {\bf q}}_a+{\dot {\bf q}}_{a-1} )\Big( E_1(\bfq_a+\bfq_{a-1}-\eta)-E_1(\bfq_a+\bfq_{a-1}+\eta) \Big)\,.
}
 \end{array}
 \eq
Finally, one should substitute the expression for ${\dot {\bf p}}_a$ from (\ref{w479}) with
\beq\label{w485}
  \begin{array}{c}
  \displaystyle{
\frac{b_1^a}{b_1^a+b_2^a}=\frac12+c{\dot\bfq}_a\,,\qquad
\frac{b_2^a}{b_1^a+b_2^a}=\frac12-c{\dot\bfq}_a\,,
}
 \end{array}
 \eq
which follows from (\ref{w481}). In this way from (\ref{w483}) one obtains:
\beq\label{w486}
  \begin{array}{c}
  \displaystyle{
\frac{4c{\ddot {\bf q}}_a}{1-4c^2{\dot {\bf q}}_a^2}=
}
\\ \ \\
  \displaystyle{
=-\frac{1}{2c}\Big(E_1(\bfq_a\!-\!\bfq_{a-1}\!-\!\eta)+E_1(\bfq_a\!+\!\bfq_{a-1}\!-\!\eta)+
E_1(\bfq_a\!-\!\bfq_{a-1}\!+\!\eta)+E_1(\bfq_a\!+\!\bfq_{a-1}\!+\!\eta)+
}
 \end{array}
 \eq
 $$
  \displaystyle{
+E_1(\bfq_a\!-\!\bfq_{a+1}\!-\!\eta)+E_1(\bfq_a\!+\!\bfq_{a+1}\!-\!\eta)+
E_1(\bfq_a\!-\!\bfq_{a+1}\!+\!\eta)+E_1(\bfq_a\!+\!\bfq_{a+1}\!+\!\eta)-4E_1(2\bfq_a)\Big)+
}
$$
$$
+{\dot\bfq}_{a+1}\Big( E_1(\bfq_a\!-\!\bfq_{a+1}\!-\!\eta)+E_1(\bfq_a\!+\!\bfq_{a+1}\!+\!\eta)
-E_1(\bfq_a\!-\!\bfq_{a+1}\!+\!\eta)-E_1(\bfq_a\!+\!\bfq_{a+1}\!-\!\eta) \Big)-
$$
$$
-{\dot\bfq}_{a-1}\Big( E_1(\bfq_a\!-\!\bfq_{a-1}\!-\!\eta)+E_1(\bfq_a\!+\!\bfq_{a-1}\!+\!\eta)
-E_1(\bfq_a\!-\!\bfq_{a-1}\!+\!\eta)-E_1(\bfq_a\!+\!\bfq_{a-1}\!-\!\eta) \Big)\,.
$$
In the special case
\beq\label{w487}
  \begin{array}{c}
  \displaystyle{
c=-\frac12
}
 \end{array}
 \eq
equations (\ref{w486}) exactly coincide with the equations (\ref{w03}) from \cite{Adler}.
Our Lax representation (\ref{w475}) differs from the one suggested in \cite{Adler}.
Presumably, there is a gauge transformation relating these two Lax representations.

\subsection{Lax pair}
\begin{predl}
Equations of motion for the elliptic Ruijsenaars-Toda chain (\ref{w478})-(\ref{w479}) or (\ref{w486})
are represented in the form of semi-discrete Zakharov-Shabat equation
\beq\label{w488}
 \begin{array}{c}
  \displaystyle{
  {{\dot {\bar L}}^a}(z)=\{H^{\hbox{\tiny{eRT}}},{\bar L}^a(z)\}={\bar L}^a(z){\bar M}^a(z)-{\bar M}^{a-1}(z){\bar L}^a(z)
 }
 \end{array}
  \eq
with the Lax matrix (\ref{w475}) and $M$-matrix
\beq\label{w489}
 \begin{array}{c}
  \displaystyle{
  {{\bar M}^a}(z)=
  \mats{{\bar M}^a_{11}(z)}
  {\displaystyle{ -\phi(z,2\bfq_a)\,\frac{1}{c}\frac{b_2^a}{b_1^a+b_2^a} }}
  {\displaystyle{ -\phi(z,-2\bfq_a)\,\frac{1}{c}\frac{b_1^a}{b_1^a+b_2^a} }}
  {{\bar M}^a_{22}(z)}\,,
 }
 \end{array}
  \eq
where
\beq\label{w490}
 \begin{array}{c}
  \displaystyle{
{\bar M}^a_{11}(z)=-\Big(\frac{1}{2c}+{\dot\bfq}_a\Big)E_1(z)+\Big(\frac{1}{2c}-{\dot\bfq}_a\Big)E_1(2\bfq_a)+
 }
 \end{array}
  \eq
$$
  \displaystyle{
+\frac{1}{4c}
\Big( E_1(\bfq_{a+1}-\bfq_a+\eta)+E_1(\bfq_{a+1}-\bfq_a-\eta)
-E_1(\bfq_{a+1}+\bfq_a+\eta)-E_1(\bfq_{a+1}+\bfq_a-\eta) \Big)+
 }
$$
$$
  \displaystyle{
+\frac12\,{\dot\bfq}_{a+1}
\Big( E_1(\bfq_{a+1}-\bfq_a-\eta)+E_1(\bfq_{a+1}+\bfq_a+\eta)
-E_1(\bfq_{a+1}+\bfq_a-\eta)-E_1(\bfq_{a+1}-\bfq_a+\eta) \Big)
 }
$$
and
\beq\label{w491}
 \begin{array}{c}
  \displaystyle{
{\bar M}^a_{22}(z)=-\Big(\frac{1}{2c}-{\dot\bfq}_a\Big)E_1(z)-\Big(\frac{1}{2c}+{\dot\bfq}_a\Big)E_1(2\bfq_a)+
 }
 \end{array}
  \eq
$$
  \displaystyle{
-\frac{1}{4c}
\Big( E_1(\bfq_{a+1}-\bfq_a+\eta)+E_1(\bfq_{a+1}-\bfq_a-\eta)
-E_1(\bfq_{a+1}+\bfq_a+\eta)-E_1(\bfq_{a+1}+\bfq_a-\eta) \Big)-
 }
$$
$$
  \displaystyle{
-\frac12\,{\dot\bfq}_{a+1}
\Big( E_1(\bfq_{a+1}-\bfq_a-\eta)+E_1(\bfq_{a+1}+\bfq_a+\eta)
-E_1(\bfq_{a+1}+\bfq_a-\eta)-E_1(\bfq_{a+1}-\bfq_a+\eta) \Big)\,.
 }
$$
\end{predl}
The proof is based on the identities (\ref{a07}), (\ref{a08}), (\ref{a09}).
%

Notice that the presented results remain valid for the trigonometric and rational limits of the
elliptic functions. In trigonometric limit $E_1(x)\rightarrow \coth(x)$,
 $\phi(z,u)\rightarrow \coth(z)+\coth(u)$, and in the rational limit
 $E_1(x)\rightarrow 1/x$,
 $\phi(z,u)\rightarrow 1/z+1/u$.

\subsection{Classical $r$-matrix structure}
Obviously, the classical $r$-matrix structure of the elliptic Ruijsenaars-Toda chain (\ref{w475})
is given by (\ref{w320})-(\ref{w327}) for $N=2$ with the identification of variables (\ref{w473}).
Let us write it down explicitly:
\begin{equation}\label{w4901}
 \begin{array}{c}
   \displaystyle{
     c\{{\bar L}_{1}^a(z),{\bar L}_2^b(w)\}=
     \delta^{ab}\Big({\bar L}_1^b(z){\bar L}_2^b(w)r^b_{12}(z,w)-r_{12}^{b-1}(z,w){\bar L}_1^b(z){\bar L}_2^b(w)+
   }
     \\ \ \\
   \displaystyle{
     +{\bar L}_1^b(z)s^{+,\,b}_{12}(z){\bar L}_2^b(w)-{\bar L}_2^b(w)s_{12}^{-,\,b}(w){\bar L}_1^b(z)\Big)+
   }
     \\ \ \\
   \displaystyle{
     +\delta^{a,\,b-1}{\bar L}_1^{b-1}(z){\bar L}_2^b(w)s_{12}^{-,\,b}(w)-
     \delta^{a,\,b+1}{\bar L}_1^{b+1}(z){\bar L}_2^b(w)s_{12}^{+,\, b+1}(z)\,,
   }
 \end{array}
\end{equation}
where
\begin{equation}\label{w4911}
    r^a_{12}(z,w)=r^a_{12}(z-w)=
    \begin{pmatrix}
        E_1(z-w)&0&0&0\\
        0&E_1(2\mathbf{q}_a)&\phi(z-w,-2\mathbf{q}_a)&0\\
        0&\phi(z-w,2\mathbf{q}_a)&-E_1(\mathbf{2q}_a)&0\\
        0&0&0&E_1(z-w)
    \end{pmatrix},
\end{equation}
\begin{equation}\label{w4921}
    s_{12}^{+,\,a}(z)=s_{12}^a(z)+u_{12}^{+,\,a}\,,
\quad
    s_{12}^{-,\,a}(w)=s_{21}^a(w)-u_{12}^{-,\,a}\,,
\end{equation}
\begin{equation}\label{w4931}
    u_{12}^{+,a}=\dfrac{1}{2}
    \begin{pmatrix}
    C&0&0&0\\
    0&D&0&0\\
    0&0&-C&0\\
    0&0&0&-D
    \end{pmatrix},
\end{equation}
\begin{equation}\label{w4941}
    u_{12}^{-,a}=\dfrac{1}{2}
    \begin{pmatrix}
    -C&0&0&0\\
    0&C&0&0\\
    0&0&-D&0\\
    0&0&0&D
    \end{pmatrix},
\end{equation}
and
\beq\label{w4951}
\begin{gathered}
    C=E_1(\mathbf{q}_{a-1}-\mathbf{q}_a+\eta)+E_1(\mathbf{q}_{a-1}+\mathbf{q}_a-\eta)\,,\\ D=E_1(\mathbf{q}_{a-1}+\mathbf{q}_a+\eta)+E_1(\mathbf{q}_{a-1}-\mathbf{q}_a-\eta)\,.
\end{gathered}
\eq
The matrices $s_{12}^a(z)$ are defined through
\begin{equation}\label{w4961}
\bar{L}_1^a(z)s_{12}^a(z)=\dfrac{1}{2}
    \begin{pmatrix}
        ''\partial_{\eta}\bar{L}^a(z)''_{11}&''\partial_{\eta}\bar{L}^a(z)''_{12}&0&0\\
        -''\partial_{\eta}\bar{L}^a(z)''_{21}&-''\partial_{\eta}\bar{L}^a(z)''_{22}&0&0\\
        0&0&-''\partial_{\eta}\bar{L}^a(z)''_{11}&-''\partial_{\eta}\bar{L}^a(z)''_{12}\\
        0&0&''\partial_{\eta}\bar{L}^a(z)''_{21}&''\partial_{\eta}\bar{L}^a(z)''_{22}
    \end{pmatrix},
\end{equation}
and
\begin{equation}\label{w4971}
\bar{L}_2^a(z)s_{21}^a(z)=\dfrac{1}{2}
    \begin{pmatrix}
        ''\partial_{\eta}\bar{L}^a(z)''_{11}&0&''\partial_{\eta}\bar{L}^a(z)''_{12}&0\\
        0&-''\partial_{\eta}\bar{L}^a(z)''_{11}&0&-''\partial_{\eta}\bar{L}^a(z)''_{12}\\
        -''\partial_{\eta}\bar{L}^a(z)''_{21}&0&-''\partial_{\eta}\bar{L}^a(z)''_{22}&0\\
        0&''\partial_{\eta}\bar{L}^a(z)''_{21}&0&''\partial_{\eta}\bar{L}^a(z)''_{22}
    \end{pmatrix},
\end{equation}
where we use the notation
\begin{equation}\label{w4981}
    ''\partial_{\eta}\bar{L}^a(z)''_{ij}=
    \bar{L}_{ij}^a(z)\Big(E_1(z+\bar{q}_i^{a-1}-\bar{q}_{j}^{a}+\eta)-E_1(\bar{q}_i^{a-1}-\bar{q}_{j}^{a}+\eta)\Big)\,.
\end{equation}

\subsection{Gauge transformation to XYZ chain}
\paragraph{$\eta$-dependent description through relativistic top.}
Following Section \ref{sec33} here we show that the Ruijsenaars-Toda chain
is gauge equivalent to the classical XYZ spin chain.
The intertwining matrix $g(z,\bfq_a)$ (\ref{w351})-(\ref{w353}) takes the following form for $N=2$:
\beq\label{w410}
  \begin{array}{c}
  \displaystyle{
g(z,\bfq_a)=
\mat{\theta_3(z-2\bfq_a|2\tau)}{\theta_3(z+2\bfq_a|2\tau)}{-\theta_2(z-2\bfq_a|2\tau)}{-\theta_2(z+2\bfq_a|2\tau)}
\mat{\frac{1}{\vth(2\bfq_a|\tau)}}{0}{0}{-\frac{1}{\vth(2\bfq_a|\tau)}}\,.
}
 \end{array}
 \eq
Here we use the Jacobi theta functions (\ref{a031}).
 The gauge transformation (\ref{w356}) then maps the monodromy matrix $T(z)$ (\ref{w103}) to
\beq\label{w411}
  \begin{array}{c}
  \displaystyle{
{\mathcal T}(z)={\mL}^1(z){\mL}^2(z)...{\mL}^n(z)
}
 \end{array}
 \eq
with
\beq\label{w412}
  \begin{array}{c}
  \displaystyle{
{\mL}^a(z)=\frac{\vth'(0)}{\vth(\eta)}\,
g(z+2\eta,\bfq_a)\,{\rm diag}(e^{\bfp_a/2c},e^{-\bfp_a/2c})\,g^{-1}(z,\bfq_a)\,.
}
 \end{array}
 \eq
Using identities (\ref{a581}) and (\ref{w6231})-(\ref{w6271}) this Lax matrix is represented in the form
\beq\label{w413}
  \begin{array}{c}
  \displaystyle{
{\mL}^a(z)=\sum\limits_{k=0}^3S^a_k\sigma_k\vf_k(z,\eta+\om_k)\,,
}
 \end{array}
 \eq
where $\sigma_k$ are the Pauli matrices (\ref{a092}), $\om_k$ are half-periods (\ref{a091})
and $\vf_k(z,\eta+\om_k)$ is a set of functions (\ref{a211}). Direct computation provides
the coefficients $S^a_k$
as functions of $\bfp_a$, $\bfq_a$ and $\eta$.
\begin{predl}
\beq\label{w414}
  \begin{array}{c}
  \displaystyle{
S_0^a=\frac12\Big(\frac{\vth(2\bfq_a-\eta)}{\vth(2\bfq_a)}\,e^{\bfp_a/2c}+
\frac{\vth(2\bfq_a+\eta)}{\vth(2\bfq_a)}\,e^{-\bfp_a/2c}\Big)\,,
}
 \end{array}
 \eq
\beq\label{w415}
  \begin{array}{c}
  \displaystyle{
S_1^a=-\frac12\frac{\theta_4(\eta)}{\vth(\eta)}
\Big(\frac{\theta_4(2\bfq_a-\eta)}{\vth(2\bfq_a)}\,e^{\bfp_a/2c}-
\frac{\theta_4(2\bfq_a+\eta)}{\vth(2\bfq_a)}\,e^{-\bfp_a/2c}\Big)\,,
}
 \end{array}
 \eq
\beq\label{w416}
  \begin{array}{c}
  \displaystyle{
S_2^a=-\frac{\imath}{2}\frac{\theta_3(\eta)}{\vth(\eta)}
\Big(\frac{\theta_3(2\bfq_a-\eta)}{\vth(2\bfq_a)}\,e^{\bfp_a/2c}-
\frac{\theta_3(2\bfq_a+\eta)}{\vth(2\bfq_a)}\,e^{-\bfp_a/2c}\Big)\,,
}
 \end{array}
 \eq
\beq\label{w417}
  \begin{array}{c}
  \displaystyle{
S_3^a=-\frac12\frac{\theta_2(\eta)}{\vth(\eta)}
\Big(\frac{\theta_2(2\bfq_a-\eta)}{\vth(2\bfq_a)}\,e^{\bfp_a/2c}-
\frac{\theta_2(2\bfq_a+\eta)}{\vth(2\bfq_a)}\,e^{-\bfp_a/2c}\Big)\,.
}
 \end{array}
 \eq
The quadratic algebra (\ref{w360}) takes the following form in the $N=2$ case.
For distinct $i,j,k\in\{1,2,3\}$
\beq\label{w418}
  \begin{array}{c}
  \displaystyle{
c\{S^a_i, S^a_j \}=\imath\varepsilon_{ijk}
I_k {S}^a_0{ S^a_k }\,,
}
\\ \ \\
  \displaystyle{
c\{S^a_0,{ S^a_i}\}=\imath\varepsilon_{ijk} (I_j-I_k)
{S^a_j} { S^a_k}\,,
}
 \end{array}
 \eq
where
\beq\label{w419}
  \begin{array}{c}
  \displaystyle{
I_k=E_1(\om_k)+ E_1(\eta)-E_1(\eta+\om_k)\,,\quad k=1,2,3\,.
}
 \end{array}
 \eq
The quadratic Poisson brackets (\ref{w418}) are generated by the classical quadratic exchange relations
(\ref{w361}) with $r$-matrix
\beq\label{w420}
  \begin{array}{c}
  \displaystyle{
r_{12}(z-w)=\frac12\,E_1(z-w)\sigma_0\otimes\sigma_0+\frac12\sum\limits_{k=1}^3\vf_k(z-w)\sigma_k\otimes\sigma_k\,,
}
 \end{array}
 \eq
where $\vf_k(z)$ is the set of functions (\ref{a27}).
\end{predl}
Formulae (\ref{w414})-(\ref{w417}) can be view as a classical analogue
for representation of quantum Sklyanin algebra \cite{Skl3} by difference operators. The Poisson brackets
(\ref{w418}) are valid for (\ref{w414})-(\ref{w417}) computed via canonical brackets (\ref{w474}).

From viewpoint of classical mechanics the expression (\ref{w413}) for any fixed $a$ is the Lax matrix of ${\rm GL}_2$ relativistic elliptic top \cite{LOZ14}
related to 8-vertex Baxter's quantum elliptic $R$-matrix \cite{Baxter}:
\beq\label{w421}
  \begin{array}{c}
  \displaystyle{
R^\eta_{12}(z-w)=\frac12\sum\limits_{k=0}^3\vf_k(z-w,\om_k+\eta)\sigma_k\otimes\sigma_k\,.
}
 \end{array}
 \eq
Namely,
\beq\label{w422}
  \begin{array}{c}
  \displaystyle{
\mL^a(z)=\tr_2\Big(R^\eta_{12}(z)S_2^a\Big)\,,\quad S_2^a=\sigma_0\otimes S^a\,.
}
 \end{array}
 \eq
This type Lax matrices give rise to a family of integrable chains \cite{DZ}.
 In the rational case the description presented here is related to 11-vertex $R$-matrix, see \cite{LOZ142}.

\paragraph{Standard description of XYZ model.} The above given description was useful for
observing relation to Ruijsenaars chain. The standard description of the XYZ model \cite{Skl4,FT}
appears in the following way. Due to relation\footnote{The relation (\ref{w423}) follows directly
from the definitions (\ref{a211}) and (\ref{a27}).} (see \cite{LOZ14})
 \beq\label{w423}
 \begin{array}{c}
  \displaystyle{
  \frac{\vf_a(z-\eta,\om_a+\eta)}{\phi(z-\eta,\eta)}=\frac{\vf_a(z,\om_a)}{\vf_a(\eta,\om_a)}
 }
 \end{array}
 \eq
we have
 \beq\label{w424}
 \begin{array}{c}
  \displaystyle{
   \frac{1}{\phi(z-\eta,\eta)}\,\mL^a(z-\eta,S^a)={\mathbb L}(z,{\mathbb S}^a)\,,
 }
 \end{array}
 \eq
 where
 \beq\label{w425}
 \begin{array}{c}
  \displaystyle{
{\mathbb L}(z,{\mathbb S}^a)=\sigma_0{\mathbb S}^a_0+\sum\limits_{k=1}^3\sigma_k\vf_k(z){\mathbb S}^a_k\,,
 }
 \end{array}
 \eq
 which is the standard expression for the Lax matrix of the classical XYZ chain \cite{Skl4,FT}. Therefore, by
 transforming the monodromy matrix ${\mathcal T}(z)$ (\ref{w411}) as
 \beq\label{w426}
 \begin{array}{c}
  \displaystyle{
{\mathbb T}(z)=\frac{1}{\phi(z-\eta,\eta)^n}{\mathcal T}(z-\eta)
 }
 \end{array}
 \eq
 we come to
 \beq\label{w427}
 \begin{array}{c}
  \displaystyle{
{\mathbb T}(z)={\mathbb L}(z,{\mathbb S}^1){\mathbb L}(z,{\mathbb S}^2)...{\mathbb L}(z,{\mathbb S}^n)\,.
 }
 \end{array}
 \eq
Due to (\ref{w423}) the elements of matrices ${\mathbb S}^a$ are simply related to the elements of matrices $S^a$:
 \beq\label{w428}
 \begin{array}{c}
  \displaystyle{
{\mathbb S}^a_0=S^a_0\,,\quad {\mathbb S}^a_k=\frac{1}{\vf_k(\eta)}\,S^a_k\,,\quad k=1,2,3\,.
 }
 \end{array}
 \eq
 From (\ref{w414})-(\ref{w417}) we get
\beq\label{w429}
  \begin{array}{c}
  \displaystyle{
{\mathbb S}_0^a=\frac12\Big(\frac{\vth(2\bfq_a-\eta)}{\vth(2\bfq_a)}\,e^{\bfp_a/2c}+
\frac{\vth(2\bfq_a+\eta)}{\vth(2\bfq_a)}\,e^{-\bfp_a/2c}\Big)\,,
}
 \end{array}
 \eq
\beq\label{w430}
  \begin{array}{c}
  \displaystyle{
{\mathbb S}_1^a=-\frac12\frac{\theta_4(0)}{\vth'(0)}
\Big(\frac{\theta_4(2\bfq_a-\eta)}{\vth(2\bfq_a)}\,e^{\bfp_a/2c}-
\frac{\theta_4(2\bfq_a+\eta)}{\vth(2\bfq_a)}\,e^{-\bfp_a/2c}\Big)\,,
}
 \end{array}
 \eq
\beq\label{w431}
  \begin{array}{c}
  \displaystyle{
{\mathbb S}_2^a=-\frac{\imath}{2}\frac{\theta_3(0)}{\vth'(0)}
\Big(\frac{\theta_3(2\bfq_a-\eta)}{\vth(2\bfq_a)}\,e^{\bfp_a/2c}-
\frac{\theta_3(2\bfq_a+\eta)}{\vth(2\bfq_a)}\,e^{-\bfp_a/2c}\Big)\,,
}
 \end{array}
 \eq
\beq\label{w432}
  \begin{array}{c}
  \displaystyle{
{\mathbb S}_3^a=-\frac12\frac{\theta_2(0)}{\vth'(0)}
\Big(\frac{\theta_2(2\bfq_a-\eta)}{\vth(2\bfq_a)}\,e^{\bfp_a/2c}-
\frac{\theta_2(2\bfq_a+\eta)}{\vth(2\bfq_a)}\,e^{-\bfp_a/2c}\Big)\,.
}
 \end{array}
 \eq
 These are the generators of the classical Sklyanin algebra \cite{Skl2}.
 \begin{predl}
 The change of variables (\ref{w429})-(\ref{w432}) provides the Poisson
 map\footnote{This statement means that the Poisson brackets for $\mathbb{S}^a(\bfp_a,\bfq_a)$
 computed through (\ref{w474}) have the form (\ref{w433}).}
  between the canonical
 Poisson structure (\ref{w474}) for the variables $\bfp_a$, $\bfq_a$ and the classical Sklyanin
 algebra for $\mathbb{S}^a$:
\beq\label{w433}
  \begin{array}{c}
  \displaystyle{
c\{\mathbb{S}^a_i,\mathbb{S}^a_j\}=\imath\varepsilon_{ijk} \mathbb{S}^a_0 \mathbb{S}^a_k\,,
}
\\ \ \\
  \displaystyle{
c\{\mathbb{S}^a_0,\mathbb{S}^a_i\}=
\imath\varepsilon_{ijk} \mathbb{S}^a_j \mathbb{S}^a_k\big(\wp(\om_j)-\wp(\om_k)\big)\,.
}
 \end{array}
 \eq
\end{predl}
The proof of these type statements is by direct computation. See e.g. \cite{MoZ}, where more general
cases are considered and proved.

The algebra (\ref{w433}) has two Casimir functions
\beq\label{w434}
  \begin{array}{c}
  \displaystyle{
{\bf C}_1^a=(\mathbb{S}^a_1)^2+(\mathbb{S}^a_2)^2+(\mathbb{S}^a_3)^2\,,
\qquad
{\bf C}_2^a=(\mathbb{S}^a_0)^2+\sum\limits_{k=1}^3 (\mathbb{S}^a_k)^2\wp(\om_k)
}
 \end{array}
 \eq
appearing from
\beq\label{w435}
  \begin{array}{c}
  \displaystyle{
\det {\mathbb L}(z,{\mathbb S}^a)={\bf C}_2^a-\wp(z){\bf C}_1^a\,.
}
 \end{array}
 \eq
Plugging (\ref{w429})-(\ref{w432}) into (\ref{w434}) one gets
\beq\label{w436}
  \begin{array}{c}
  \displaystyle{
{\bf C}_1^a=\Big(\frac{\vth(\eta)}{\vth'(0)}\Big)^2\,,
\qquad
{\bf C}_2^a=\wp(\eta)\Big(\frac{\vth(\eta)}{\vth'(0)}\Big)^2\,.
}
 \end{array}
 \eq
Finally, let us remark that in the continuous non-relativistic limit the XYZ chain turns into the classical Landau-Lifshitz magnet (1+1 field theory) \cite{FT}. An analogue of the gauge transformation (\ref{w356}) relates it
with the 1+1 field version of 2-body Calogero-Moser model, and the formulae similar to (\ref{w414})-(\ref{w417}) can be derived in the field case as well \cite{AtZ}.

\paragraph{Hamiltonian.} The Hamiltonian of the Ruijsenaars (and the Ruijsenaars-Toda) chain is computed
as logarithm of trace of the coefficient behind the highest order pole in $z$ (that is, behind $1/z^n$)
 for the monodromy
matrix $T(z)$ (\ref{w105}).
The Hamiltonian for the XYZ chain can be defined in the same way since
\beq\label{w437}
  \begin{array}{c}
  \displaystyle{
\res\limits_{z=0}\mL^a(z)=S^a={\mathbb L}(\eta,{\mathbb S}^a)\,,
}
 \end{array}
 \eq
which follows from (\ref{w428}). Then
\beq\label{w438}
  \begin{array}{c}
  \displaystyle{
\det{\mathbb L}(\eta,{\mathbb S}^a)=\det S^a=0\,.
}
 \end{array}
 \eq
The latter is valid due to (\ref{w436}) by plugging it into (\ref{w435}) together with $z=\eta$.
Therefore, $S^a$ are degenerated matrices, and they are represented in the form
\beq\label{w439}
  \begin{array}{c}
  \displaystyle{
S^a=\xi^a\otimes\psi^a\,,
}
 \end{array}
 \eq
where $\xi^a$ are 2-dimensional column vectors, while $\psi^a$ are  2-dimensional row vectors.
Then
\beq\label{w440}
  \begin{array}{c}
  \displaystyle{
H=c\log\tr(S^1S^2...S^n)=c\log\tr(\xi^1\otimes\psi^1...\xi^n\otimes\psi^n)=c\sum\limits_{a=1}^n \log h_{a,a+1}\,,
}
 \end{array}
 \eq
where $h_{a-1,a}$ are scalar products
\beq\label{w441}
  \begin{array}{c}
  \displaystyle{
h_{a-1,a}=(\psi^{a-1},\xi^a)\,.
}
 \end{array}
 \eq
The vectors $\xi^a$ and covectors $\psi^a$ are defined up to multiplication $\xi^a\rightarrow\lambda_a\xi^a$,
$\psi^a\rightarrow \lambda_a^{-1}\psi^a$. This allows to identify $h_{a-1,a}$ (\ref{w441}) with $h^{\hbox{\tiny{eRT}}}_{a-1,a}$.

Unfortunately, the Poisson structure is known in terms of matrices $S^a$ (\ref{w418}) only, but it is unknown
in terms of $\xi^a$, $\psi^a$. However, it is possible to find $M$-matrices $M^a(z)$ satisfying the
semi-discrete Zakharov-Shabat equation (\ref{w109}) and the corresponding equations of motion reproduce the
Landau-Lifshitz magnet in the continuous limit \cite{DZ}.

Let us also recall that in the standard description \cite{Skl4,FT} the Hamiltonian of the XYZ chain
is given as
\beq\label{w442}
  \begin{array}{c}
  \displaystyle{
H^{\hbox{\tiny{XYZ}}}=c\sum\limits_{a=1}^n\log\tr\Big( {\mathbb L}(\eta,{\mathbb S}^a){\mathbb L}(\eta,{\mathbb S}^{a+1})\Big)=
c\sum\limits_{a=1}^n\log\tr\Big(S^aS^{a+1}\Big)=
}
\\
  \displaystyle{
=c\sum\limits_{a=1}^n\Big(\log(\psi^{a},\xi^{a+1})+\log(\psi^{a+1},\xi^{a})\Big)\,.
}
 \end{array}
 \eq
The second term in the last line can be obtained by considering
\beq\label{w443}
  \begin{array}{c}
  \displaystyle{
\log\tr \Big({\mathbb T}(z)\Big)+\log\tr\Big(\det{\mathbb T}(z) {\mathbb T}^{-1}(-z)\Big)
}
 \end{array}
 \eq
as generating function of Hamiltonians (instead of $\log\tr {\mathbb T}(z)$ providing (\ref{w440})).
Due to oddness $\vf_\al(-z)=-\vf_\al(z)$ we have
\beq\label{w444}
  \begin{array}{c}
  \displaystyle{
\det{\mathbb T}(z) {\mathbb T}^{-1}(-z)={\mathbb L}^n(z){\mathbb L}^{n-1}(z)...{\mathbb L}^1(z)\,.
}
 \end{array}
 \eq
This is why we obtain (\ref{w442}).

From viewpoint of the elliptic Ruijsenaars-Toda chain,
the first contribution to (\ref{w442}) equals
$cH^{\hbox{\tiny{eRT}}}$. Denote the expression $h^{\hbox{\tiny{eRT}}}_{a-1,a}$
(\ref{w477}) as $h^{\hbox{\tiny{eRT}}}_{a-1,a}(\bfq_{a-1},\bfq_a,\bfp_a,\eta,c)$.
Then
\beq\label{w445}
  \begin{array}{c}
  \displaystyle{
H^{\hbox{\tiny{XYZ}}}=c\sum\limits_{a=1}^n\Big(\log h^{\hbox{\tiny{eRT}}}_{a-1,a}(\bfq_{a-1},\bfq_a,\bfp_a,\eta,c)
+\log h^{\hbox{\tiny{eRT}}}_{a+1,a}(\bfq_{a+1},\bfq_a,\bfp_a,-\eta,-c)\Big)\,,
}
 \end{array}
 \eq
where the sum of the first terms provides $c H^{\hbox{\tiny{eRT}}}$.

\subsection{How to introduce more parameters $\eta_a$}\label{sec44}
In \cite{Adler} the elliptic Ruijsenaars-Toda chain (\ref{w03}) was defined not for a single parameter
$\eta$ but for a set of parameters $\eta_a$, $a=1,...,n$. In order to explain how to introduce more parameters
let us return back to the $\eta$-dependent description of XYZ model (\ref{w411})-(\ref{w413}).
Consider the following natural generalization of the monodromy matrix (\ref{w411}):
\beq\label{w4111}
  \begin{array}{c}
  \displaystyle{
{\mathcal T}(z)={\mL}^1(z,\eta_1){\mL}^2(z,\eta_2)...{\mL}^n(z,\eta_n)
}
 \end{array}
 \eq
with
\beq\label{w4121}
  \begin{array}{c}
  \displaystyle{
{\mL}^a(z,\eta_a)=\frac{\vth'(0)}{\vth(\eta_a)}\,
g(z+2\eta_a,\bfq_a)\,{\rm diag}(e^{\bfp_a/2c},e^{-\bfp_a/2c})\,g^{-1}(z,\bfq_a)\,.
}
 \end{array}
 \eq
Then the Lax matrix has the form
\beq\label{w4131}
  \begin{array}{c}
  \displaystyle{
{\mL}^a(z,\eta_a)=\sum\limits_{k=0}^3S^a_k\sigma_k\vf_k(z,\eta_a+\om_k)\,.
}
 \end{array}
 \eq
The model with the monodromy matrix (\ref{w4111}) is integrable since each ${\mL}^a(z,\eta_a)$ satisfies
the quadratic $r$-matrix structure (\ref{w361}) with the same elliptic $r$-matrix (\ref{w420}).
At each site we have a straightforward generalization of (\ref{w414}):
\beq\label{w4141}
  \begin{array}{c}
  \displaystyle{
S_0^a=\frac12\Big(\frac{\vth(2\bfq_a-\eta_a)}{\vth(2\bfq_a)}\,e^{\bfp_a/2c}+
\frac{\vth(2\bfq_a+\eta_a)}{\vth(2\bfq_a)}\,e^{-\bfp_a/2c}\Big)\,,
}
 \end{array}
 \eq
\beq\label{w4151}
  \begin{array}{c}
  \displaystyle{
S_1^a=-\frac12\frac{\theta_4(\eta_a)}{\vth(\eta_a)}
\Big(\frac{\theta_4(2\bfq_a-\eta_a)}{\vth(2\bfq_a)}\,e^{\bfp_a/2c}-
\frac{\theta_4(2\bfq_a+\eta_a)}{\vth(2\bfq_a)}\,e^{-\bfp_a/2c}\Big)\,,
}
 \end{array}
 \eq
\beq\label{w4161}
  \begin{array}{c}
  \displaystyle{
S_2^a=-\frac{\imath}{2}\frac{\theta_3(\eta_a)}{\vth(\eta_a)}
\Big(\frac{\theta_3(2\bfq_a-\eta_a)}{\vth(2\bfq_a)}\,e^{\bfp_a/2c}-
\frac{\theta_3(2\bfq_a+\eta_a)}{\vth(2\bfq_a)}\,e^{-\bfp_a/2c}\Big)\,,
}
 \end{array}
 \eq
\beq\label{w4171}
  \begin{array}{c}
  \displaystyle{
S_3^a=-\frac12\frac{\theta_2(\eta_a)}{\vth(\eta_a)}
\Big(\frac{\theta_2(2\bfq_a-\eta_a)}{\vth(2\bfq_a)}\,e^{\bfp_a/2c}-
\frac{\theta_2(2\bfq_a+\eta_a)}{\vth(2\bfq_a)}\,e^{-\bfp_a/2c}\Big)\,,
}
 \end{array}
 \eq
and the Poisson brackets for the corresponding Sklyanin algebras are as follows:
\beq\label{w4181}
  \begin{array}{c}
  \displaystyle{
c\{S^a_i, S^a_j \}=\imath\varepsilon_{ijk}
I^a_k {S}^a_0{ S^a_k }\,,
}
\\ \ \\
  \displaystyle{
c\{S^a_0,{ S^a_i}\}=\imath\varepsilon_{ijk} (I^a_j-I^a_k)
{S^a_j} { S^a_k}\,,
}
 \end{array}
 \eq
where
\beq\label{w4191}
  \begin{array}{c}
  \displaystyle{
I^a_k=E_1(\eta_a)+ E_1(\om_k)-E_1(\eta_a+\om_k)\,,\quad k=1,2,3\,.
}
 \end{array}
 \eq

It is interesting to notice that the above given construction is problematic
in the standard description. Indeed, due to (\ref{w424}) in the standard description
we obtain the monodromy matrix
\beq\label{w4451}
  \begin{array}{c}
  \displaystyle{
{\mathbb  T}(z)={\mathbb L}^1(z+\eta_1){\mathbb L}^2(z+\eta_2)...{\mathbb L}^n(z+\eta_n)\,,
}
 \end{array}
 \eq
that is parameters $\eta_a$ play the role of inhomogeneous parameters. At the same time it is known that
one should consider a homogeneous model in order to have a local interaction (of neighbour sites only)
because one needs existence of a point $z=z_*$, where all ${\mathbb L}^a(z_*+\eta_a)$ are degenerated simultaneously.
On the other hand, in the $\eta$-dependent description (\ref{w4111})-(\ref{w4131}) all $S^a=\res\limits_{z=0}\mL^a(z)$
are degenerated. It happens due to factorization (\ref{w4121}) because $\res\limits_{z=0}g^{-1}(z,{\bar q}^a)$
is a rank 1 matrix, see \cite{ZZ}. In our case the Casimirs are not generic: $C_1^a=(\vth(\eta_a)^2/\vth'(0)^2)$, $C_2^a=\wp(\eta_a)C_1^a$, and $\det{\mathbb L}^a(z+\eta_a)=C_1^a(\wp(\eta_a)-\wp(z+\eta_a))$.
For this reason $z=z_*=0$ exists for all $a$ simultaneously.

Next, by making the inverse of the gauge transformation (\ref{w3562})
\beq\label{w446}
  \begin{array}{c}
  \displaystyle{
{\mathcal T}(z)\rightarrow {\bar T}(z)=g^{-1}(z,{\bar q}^{n}){\mathcal T}(z)g(z,{\bar q}^{n})
}
 \end{array}
 \eq
we come to the elliptic Ruijsenaars-Toda chain with a set of parameters $\eta_a$.
This is a direct generalization of (\ref{w475})-(\ref{w476}):
\beq\label{w4752}
  \begin{array}{c}
  \displaystyle{
{\bar L}^a(z)=\mat{\phi(z,\bfq_{a-1}-\bfq_a+\eta_a)b^a_1}{\phi(z,\bfq_{a-1}+\bfq_a+\eta_a)b^a_2}
{\phi(z,-\bfq_{a-1}-\bfq_a+\eta_a)b^a_1}{\phi(z,-\bfq_{a-1}+\bfq_a+\eta_a)b^a_2}\,,
}
 \end{array}
 \eq
where
\beq\label{w4762}
  \begin{array}{c}
  \displaystyle{
b^a_1=\frac{\vth(\bfq_a-\bfq_{a-1}-\eta_a)\vth(\bfq_a+\bfq_{a-1}-\eta_a)}{\vth(-\eta_a)\vth(2\bfq_a)}
\exp\Big(\frac{\bfp_a}{2c}\Big)\,,
}
\\ \ \\
  \displaystyle{
b^a_2=-\frac{\vth(\bfq_a-\bfq_{a-1}+\eta_a)\vth(\bfq_a+\bfq_{a-1}+\eta_a)}{\vth(-\eta_a)\vth(2\bfq_a)}
\exp\Big(-\frac{\bfp_a}{2c}\Big)\,.
}
 \end{array}
 \eq
The rest of description of this model repeats the one for a single $\eta$ (\ref{w477})-(\ref{w486}).
Finally, we come to the equations of motion
\beq\label{w4861}
  \begin{array}{c}
  \displaystyle{
\frac{4c{\ddot {\bf q}}_a}{1-4c^2{\dot {\bf q}}_a^2}=
}
\\ \ \\
  \displaystyle{
=-\frac{1}{2c}\Big(E_1(\bfq_a\!-\!\bfq_{a-1}\!-\!\eta_a)+E_1(\bfq_a\!+\!\bfq_{a-1}\!-\!\eta_a)+
E_1(\bfq_a\!-\!\bfq_{a-1}\!+\!\eta_a)+E_1(\bfq_a\!+\!\bfq_{a-1}\!+\!\eta_a)+
}
 \end{array}
 \eq
 $$
  \displaystyle{
+E_1(\bfq_a\!-\!\bfq_{a+1}\!-\!\eta_{a+1})+E_1(\bfq_a\!+\!\bfq_{a+1}\!-\!\eta_{a+1})+
E_1(\bfq_a\!-\!\bfq_{a+1}\!+\!\eta_{a+1})+E_1(\bfq_a\!+\!\bfq_{a+1}\!+\!\eta_{a+1})-
}
$$
$$
  \displaystyle{
-4E_1(2\bfq_a)\Big)+
}
$$
$$
+{\dot\bfq}_{a+1}\Big( E_1(\bfq_a\!-\!\bfq_{a+1}\!-\!\eta_{a+1})+E_1(\bfq_a\!+\!\bfq_{a+1}\!+\!\eta_{a+1})
-E_1(\bfq_a\!-\!\bfq_{a+1}\!+\!\eta_{a+1})-E_1(\bfq_a\!+\!\bfq_{a+1}\!-\!\eta_{a+1}) \Big)-
$$
$$
-{\dot\bfq}_{a-1}\Big( E_1(\bfq_a\!-\!\bfq_{a-1}\!-\!\eta_a)+E_1(\bfq_a\!+\!\bfq_{a-1}\!+\!\eta_a)
-E_1(\bfq_a\!-\!\bfq_{a-1}\!+\!\eta_a)-E_1(\bfq_a\!+\!\bfq_{a-1}\!-\!\eta_a) \Big)\,.
$$
For $c=-1/2$ these are the equations which were introduced in \cite{Adler}.

\section{Elliptic Toda chain}\label{sec6}
\setcounter{equation}{0}
The elliptic Toda chain can be considered as particular case of the Ruijsenaars-Toda chain in the limiting case
$\eta=0$ but one should previously proceed to modified Lax matrices. This is what we begin with in this Section.

\subsection{Modified Lax matrices for the Ruijsenaars chain}
Following \cite{ZZ} introduce the modified Lax matrices a follows:
 \beq\label{w501}
 \begin{array}{c}
  \displaystyle{
 {L'}^a(z)=\frac{1}{{\bar h}_{a-1,a}}\,{\bar L}^a(z)
 }
 \end{array}
 \eq
with the Lax matrices ${\bar L}^a(z)$ (\ref{w303}). Explicitly, we have
 \beq\label{w502}
 \begin{array}{c}
  \displaystyle{
 {L'}^a_{ij}(z)=\frac{1}{{\bar h}_{a-1,a}}\,{\bar L}^a_{ij}(z)=
 \phi(z,{\bar q}^{a-1}_i-{\bar q}^{a}_j+\eta) \frac{b^a_{j}}{{\bar h}_{a-1,a}}=
 \phi(z,{\bar q}^{a-1}_i-{\bar q}^{a}_j+\eta) \frac{b^a_{j}}{\sum\limits_{k=1}^Nb_k^a}
 }
 \end{array}
 \eq
for $a=1,...,n$, $i,j=1,...,N$. The meaning of this transformation is to make the Lax matrix to
depend on ${\dot q}^a_j$ (\ref{w4031}) instead of $b^a_j$.
The corresponding monodromy matrix takes the form
 \beq\label{w503}
 \begin{array}{c}
  \displaystyle{
{T'}(z)={L'}^1(z)...{L'}^n(z)=\frac{1}{{\bar h}_{1,2}{\bar h}_{2,3}...{\bar h}_{n,1}}\,{\bar T}(z)
={\bar T}(z)e^{-{\bar H}/c}\,,
 }
 \end{array}
 \eq
where ${\bar T}(z)$ is the monodromy matrix (\ref{w342}), and $\bar H$ is the Hamiltonian generated
by ${\bar T}(z)$. For this reason the Poisson commutativity
\beq\label{w504}
  \begin{array}{c}
  \displaystyle{
\{\tr \left({T'}^k(z)\right),\tr \left({ T'}^l(w)\right)\} =0
}
 \end{array}
 \eq
follows from (\ref{w342}).


\subsection{Hamiltonian and Lax pair}
Consider the Lax matrix of the Ruijsenaars-Toda chain (\ref{w475}). Then the modified Lax matrix is as follows:
\beq\label{w510}
  \begin{array}{c}
  \displaystyle{
{L'}^a(z)=\mat{\phi(z,\bfq_{a-1}-\bfq_a+\eta)
\displaystyle{\frac{b^a_1}{b^a_1+b^a_2}}}{\phi(z,\bfq_{a-1}+\bfq_a+\eta)\displaystyle{\frac{b^a_2}{b^a_1+b^a_2}}}
{\phi(z,-\bfq_{a-1}-\bfq_a+\eta)\displaystyle{\frac{b^a_1}{b^a_1+b^a_2}}}
{\phi(z,-\bfq_{a-1}+\bfq_a+\eta)\displaystyle{\frac{b^a_2}{b^a_1+b^a_2}}}\,,
}
 \end{array}
 \eq
where $b_1^a$ and $b_2^a$ are the functions (\ref{w476}). The limit to $\eta=0$ is well defined
for the ratios $\frac{b^a_1}{b^a_1+b^a_2}$ and $\frac{b^a_2}{b^a_1+b^a_2}$
since all theta functions are cancelled out:
\beq\label{w511}
  \begin{array}{c}
  \displaystyle{
\frac{b^a_1}{b^a_1+b^a_2}\Big|_{\eta=0}=
\frac{\exp\Big(\frac{\bfp_a}{2c}\Big)}{\exp\Big(\frac{\bfp_a}{2c}\Big)-\exp\Big(-\frac{\bfp_a}{2c}\Big)}\,,
}
\\ \ \\
  \displaystyle{
\frac{b^a_2}{b^a_1+b^a_2}\Big|_{\eta=0}=-
\frac{\exp\Big(-\frac{\bfp_a}{2c}\Big)}{\exp\Big(\frac{\bfp_a}{2c}\Big)-\exp\Big(-\frac{\bfp_a}{2c}\Big)}\,.
}
 \end{array}
 \eq
Therefore, for $\eta=0$ we obtain
\beq\label{w512}
  \begin{array}{c}
  \displaystyle{
{\bf L}^a(z)=\frac12\mat{\phi(z,\bfq_{a-1}-\bfq_a)
\displaystyle{\frac{e^{\bfp_a/2c}}{\sinh(\bfp_a/2c)}}}
{-\phi(z,\bfq_{a-1}+\bfq_a)\displaystyle{\frac{e^{-\bfp_a/2c}}{\sinh(\bfp_a/2c)}}}
{\phi(z,-\bfq_{a-1}-\bfq_a)\displaystyle{\frac{e^{\bfp_a/2c}}{\sinh(\bfp_a/2c)}}}
{-\phi(z,-\bfq_{a-1}+\bfq_a)\displaystyle{\frac{e^{-\bfp_a/2c}}{\sinh(\bfp_a/2c)}}}\,.
}
 \end{array}
 \eq
Due to (\ref{a19})
\beq\label{w513}
  \begin{array}{c}
  \displaystyle{
\det{\bf L}^a(z)=\frac14\frac{1}{\sinh^2(\bfp_a/2c)}\Big(\wp(\bfq_{a-1}-\bfq_a)-\wp(\bfq_{a-1}+\bfq_a)\Big)\,.
}
 \end{array}
 \eq
Then for the monodromy matrix
\beq\label{w514}
  \begin{array}{c}
  \displaystyle{
{\bf T}(z)={\bf L}^1(z){\bf L}^2(z)...{\bf L}^n(z)
}
 \end{array}
 \eq
we have
\beq\label{w515}
  \begin{array}{c}
  \displaystyle{
\log\det{\bf T}(z)=\sum\limits_{a=1}^n\Big(\log\frac{1}{\sinh^2(\bfp_a/2c)}+
\log\Big(\wp(\bfq_{a-1}-\bfq_a)-\wp(\bfq_{a-1}+\bfq_a)\Big)\Big)-\log(4^n)=
}
\\ \ \\
  \displaystyle{
  =-2H^{\hbox{\tiny{eToda}}}-\log(4^n)\,,
  }
 \end{array}
 \eq
which is independent of $z$, and $H^{\hbox{\tiny{eToda}}}$ is the Hamiltonian (\ref{w01}).
Notice that for non-modified monodromy matrix ${\bar T}(z)={\bar L}^1(z){\bar L}^2(z)...{\bar L}^n(z)$
its determinant $\det{\bar T}(z)$ is a Casimir function. The non-trivial expression (\ref{w515})
came from the factor $e^{-{\bar H}/c}$ in (\ref{w503}).

Due to (\ref{a053}) equations of motion take the form:
\beq\label{w516}
  \begin{array}{c}
  \displaystyle{
{\dot \bfq}_a=\p_{\bfp_a}H^{\hbox{\tiny{eToda}}}=\frac{1}{2c}\frac{b_1^a-b_2^a}{b_1^a+b_2^a}
=\frac{1}{2c}\coth\Big(\frac{\bfp_a}{2c}\Big)\,,
}
 \end{array}
 \eq
\beq\label{w517}
  \begin{array}{c}
  \displaystyle{
{\dot \bfp}_a=-\p_{\bfq_a}H^{\hbox{\tiny{eToda}}}=
E_1(2\bfq_a)-E_1(\bfq_a-\bfq_{a-1})-E_1(\bfq_a+\bfq_{a-1})+
}
\\ \ \\
  \displaystyle{
+E_1(2\bfq_a)-E_1(\bfq_a-\bfq_{a+1})-E_1(\bfq_a+\bfq_{a+1})
}
 \end{array}
 \eq
or in the Newtonian form
  \beq\label{w5171}
  \begin{array}{r}
  \displaystyle{
\frac{{4c^2\ddot {\bf q}}_a}{4c^2{\dot {\bf q}}_a^2-1}=
E_1(\bfq_a-\bfq_{a-1})+E_1(\bfq_a+\bfq_{a-1})-E_1(2\bfq_a)+
}
\\ \ \\
  \displaystyle{
+E_1(\bfq_a-\bfq_{a+1})+E_1(\bfq_a+\bfq_{a+1})-E_1(2\bfq_a)\,,
  }
 \end{array}
 \eq
which turn into (\ref{w02}) for $c=-1/2$.
\begin{predl}
Equations of motion (\ref{w5171}) are generated by the Zakharov-Shabat equation
\beq\label{w518}
 \begin{array}{c}
  \displaystyle{
  {{\dot {\bf L}}^a}(z)=\{H^{\hbox{\tiny{eToda}}},{\bf L}^a(z)\}=
  {\bf L}^a(z){\bf M}^a(z)-{\bf M}^{a-1}(z){\bf L}^a(z)
 }
 \end{array}
  \eq
with the Lax matrix (\ref{w512})
\beq\label{w5181}
  \begin{array}{c}
  \displaystyle{
{\bf L}^a(z)=c\mat{\phi(z,\bfq_{a-1}-\bfq_a)
\displaystyle{\Big(\frac{1}{2c}+{\dot\bfq}_a\Big)}}
{\phi(z,\bfq_{a-1}+\bfq_a)\displaystyle{\Big(\frac{1}{2c}-{\dot\bfq}_a\Big)}}
{\phi(z,-\bfq_{a-1}-\bfq_a)\displaystyle{\Big(\frac{1}{2c}+{\dot\bfq}_a\Big)}}
{\phi(z,-\bfq_{a-1}+\bfq_a)\displaystyle{\Big(\frac{1}{2c}-{\dot\bfq}_a\Big)}}
}
 \end{array}
 \eq
 and the following $M$-matrix:
\beq\label{w530}
 \begin{array}{c}
  \displaystyle{
  {{\bf M}^a}(z)=
  \mats{{\bf M}^a_{11}(z)}
  {\displaystyle{ -\phi(z,2\bfq_a)\,\frac{1}{c}\frac{b_2^a}{b_1^a+b_2^a} }}
  {\displaystyle{ -\phi(z,-2\bfq_a)\,\frac{1}{c}\frac{b_1^a}{b_1^a+b_2^a} }}
  {{\bf M}^a_{22}(z)}
 }
 \end{array}
  \eq
or
\beq\label{w5301}
 \begin{array}{c}
  \displaystyle{
  {{\bf M}^a}(z)=
  \mats{{\bf M}^a_{11}(z)}
  {\displaystyle{ -\phi(z,2\bfq_a)\Big(\frac{1}{2c}-{\dot\bfq}_a\Big) }}
  {\displaystyle{ -\phi(z,-2\bfq_a)\Big(\frac{1}{2c}+{\dot\bfq}_a\Big) }}
  {{\bf M}^a_{22}(z)}\,,
 }
 \end{array}
  \eq
where
\beq\label{w531}
 \begin{array}{c}
  \displaystyle{
{\bf M}^a_{11}(z)=-\Big(\frac{1}{2c}+{\dot\bfq}_a\Big)E_1(z)+\Big(\frac{1}{2c}-{\dot\bfq}_a\Big)
\Big(E_1(2\bfq_a)-E_1(\bfq_a-\bfq_{a+1})-E_1(\bfq_a+\bfq_{a+1})\Big)\,,
 }
 \end{array}
  \eq
\beq\label{w532}
 \begin{array}{c}
  \displaystyle{
{\bf M}^a_{22}(z)=-\Big(\frac{1}{2c}-{\dot\bfq}_a\Big)E_1(z)
-\Big(\frac{1}{2c}+{\dot\bfq}_a\Big)\Big(E_1(2\bfq_a)-E_1(\bfq_a-\bfq_{a+1})-E_1(\bfq_a+\bfq_{a+1})\Big)
 }
 \end{array}
  \eq
  and
\beq\label{w533}
 \begin{array}{c}
  \displaystyle{
\frac{1}{c}\frac{b_1^a}{b_1^a+b_2^a}=\frac{e^{\bfp_a/2c}}{2c\sinh(\bfp_a/2c)}=\frac{1}{2c}+{\dot\bfq}_a\,,
 }
 \end{array}
  \eq
\beq\label{w534}
 \begin{array}{c}
  \displaystyle{
\frac{1}{c}\frac{b_2^a}{b_1^a+b_2^a}=-\frac{e^{-\bfp_a/2c}}{2c\sinh(\bfp_a/2c)}=\frac{1}{2c}-{\dot\bfq}_a\,.
 }
 \end{array}
  \eq
  \end{predl}
  The proof of (\ref{w518})-(\ref{w530}) is based on (\ref{a09}).

  The Lax matrix in the form (\ref{w5181}) with $c=-1/2$ was proposed in \cite{KrToda} (up to some simple transformations).

\subsection{Classical $r$-matrix structure}
The simplest way to obtain the classical $r$-matrix structure for elliptic Toda chain is to
use the results (\ref{w4901})-(\ref{w4981}) for the elliptic Ruijsenaars-Toda chain.
Indeed, since the $r$-matrix structure is quadratic, we may remove the factor $\vth(-\eta)$ in the
definition of the Lax matrices (\ref{w475}). Then the limit to $\eta=0$ is well-defined:
\beq\label{w4751}
  \begin{array}{c}
  \displaystyle{
{\bar {\bf L}}^a(z)=\mat{\phi(z,\bfq_{a-1}-\bfq_a)b^a_1}{\phi(z,\bfq_{a-1}+\bfq_a)b^a_2}
{\phi(z,-\bfq_{a-1}-\bfq_a)b^a_1}{\phi(z,-\bfq_{a-1}+\bfq_a)b^a_2}\,,
}
 \end{array}
 \eq
with
\beq\label{w4761}
  \begin{array}{c}
  \displaystyle{
b^a_1=\frac{\vth(\bfq_a-\bfq_{a-1})\vth(\bfq_a+\bfq_{a-1})}{\vth(2\bfq_a)}
\exp\Big(\frac{\bfp_a}{2c}\Big)\,,
}
\\ \ \\
  \displaystyle{
b^a_2=-\frac{\vth(\bfq_a-\bfq_{a-1})\vth(\bfq_a+\bfq_{a-1})}{\vth(2\bfq_a)}
\exp\Big(-\frac{\bfp_a}{2c}\Big)\,.
}
 \end{array}
 \eq
The limit (to $\eta=0$) of all components entering (\ref{w4901})-(\ref{w4981}) is also
well-defined. In this way we obtain the final answer (which is just (\ref{w4901})-(\ref{w4981}) with $\eta=0$).

In the Lax representation (\ref{w4751}) the Hamiltonian (\ref{w01}) appears from (\ref{w477}).
Consider
\beq\label{w4765}
  \begin{array}{c}
  \displaystyle{
H_0=\prod\limits_{a=1}^n(b_1^a+b_2^a)=
2^n\prod\limits_{a=1}^n \frac{\vth(\bfq_a-\bfq_{a-1})\vth(\bfq_a+\bfq_{a-1})}{\vth(2\bfq_a)}\sinh\big(\frac{\bfp_a}{2c}\big)=
}
\\
  \displaystyle{
=2^n\prod\limits_{a=1}^n \frac{\vth(\bfq_a-\bfq_{a-1})\vth(\bfq_a+\bfq_{a-1})}{\sqrt{\vth(2\bfq_a)\vth(2\bfq_{a-1})}}
\sinh\big(\frac{\bfp_a}{2c}\big)
}
 \end{array}
 \eq
Due to (\ref{a19}) we have
\beq\label{w4766}
  \begin{array}{c}
  \displaystyle{
\wp(\bfq_a-\bfq_{a-1})-\wp(\bfq_a+\bfq_{a-1})=
-\phi(\bfq_a-\bfq_{a-1},\bfq_a+\bfq_{a-1})\phi(\bfq_{a-1}-\bfq_{a},\bfq_a+\bfq_{a-1})=
}
\\ \ \\
  \displaystyle{
=\frac{\vth'(0)^2\vth(2\bfq_a)\vth(2\bfq_{a-1})}{\vth(\bfq_a-\bfq_{a-1})^2\vth(\bfq_a+\bfq_{a-1})^2}
}
 \end{array}
 \eq
Therefore, the Hamiltonian of the elliptic Toda chain (\ref{w01}) equals $\log H_0-n\log(2\vth'(0))$.

In order to get the $r$-matrix structure for the Lax matrix (\ref{w512}) we need to
explain how proceeding to the modified Lax matrices (\ref{w502}) effects the result.
We have
\begin{equation}\label{531}
    \begin{gathered}
        \left\{L_1'^a(z),L_2'^b(w)\right\}=
        \dfrac{1}{\bar{h}_{a-1,a}\bar{h}_{b-1,b}}
        \Big(\left\{\bar{L}_1^a(z),\bar{L}_2^b(w)\right\}-
        L'^a(z)\otimes \left\{\bar{h}_{a-1,a},\bar{L}^b(w)\right\}-
        \\
        -\left\{\bar{L}^a(z),\bar{h}_{b-1,b}\right\}\otimes L'^b(w)+L_1'^a(z)L_2'^b(w)\left\{\bar{h}_{a-1,a},\bar{h}_{b-1,b}\right\}\Big).
    \end{gathered}
\end{equation}
Direct calculation of additional terms can be performed.
Below we present an answer obtained by direct computation instead of using (\ref{531}). Notice that the resultant $r$-matrix structure depends on momenta in contrast to (\ref{w4901}) for the (non-modified) Lax matrix (\ref{w475}).

Let us write the Lax matrix~\eqref{w512} in the following form:
\begin{equation}\label{w677}
    \begin{gathered}
        \mathbf{L}^a(z) = \Phi(z,\mathbf{q}_{a-1},\mathbf{q}_a)\mathbf{d}(\mathbf{p}_a)\,,
    \end{gathered}
\end{equation}
where
\begin{equation}
    \begin{gathered}
        \Phi(z,\mathbf{q}_{a-1},\mathbf{q}_a)=
        \begin{pmatrix}
            \phi(z,\mathbf{q}_{a-1}-\mathbf{q}_a)&\phi(z,\mathbf{q}_{a-1}+\mathbf{q}_a)\\ \ \\
            \phi(z,-\mathbf{q}_{a-1}-\mathbf{q}_a)&\phi(z,-\mathbf{q}_{a-1}+\mathbf{q}_a)
        \end{pmatrix}
    \end{gathered}
\end{equation}
and
\begin{equation}
\begin{gathered}
      \mathbf{d}(\mathbf{p}_a)=\text{diag}\big(d(-\mathbf{p}_a),d(\mathbf{p}_a)\big)
\end{gathered}
\end{equation}
with
$$
d(\mathbf{p}_a)=\dfrac{1}{1-\exp\left(\dfrac{\mathbf{p}_a}{c}\right)}\,.
$$
We will use the following notation:
\begin{equation}
    \mathbf{d}_1(\mathbf{p}_a) = d(-\mathbf{p_a}); \quad \mathbf{d}_2(\mathbf{p}_a) = d(\mathbf{p}_a)\,.
\end{equation}
Using straightforward calculation the Poisson brackets can be written in the following form:
\begin{equation}
    \begin{gathered}
        c\left\{\mathbf{L}_1^a(z),\mathbf{L}_2^b(w)\right\} = \delta^{ab}\Bigg[\sum_{i,j,k,l=1}^2\mathbf{L}_{ij}^a(z)\mathbf{L}_{kl}^a(w)\times\\
        \times (-1)^{j+l}\Big(\mathbf{d}_{l-1}(\mathbf{p}_a)F_{ij}^a(z)
        -\mathbf{d}_{j-1}(\mathbf{p}_a)F_{kl}^a(w)\Big)E_{ij}\otimes E_{kl}
        \Bigg]-
        \\
        -\delta^{a-1,b}\Bigg[\sum_{i,j,k,l=1}^2\mathbf{L}^a_{ij}(z)\mathbf{L}_{kl}^{a-1}(w)(-1)^{i+l}
        \mathbf{d}_{l-1}(\mathbf{p}_{a-1})F_{ij}^a(z)E_{ij}\otimes E_{kl}\Bigg]+
        \\
        +\delta^{a,b-1}\Bigg[\sum_{i,j,k,l=1}^2\mathbf{L}^a_{ij}(z)\mathbf{L}_{kl}^{a+1}(w)(-1)^{k+j}
        \mathbf{d}_{j-1}(\mathbf{p}_a)F_{kl}^{a+1}(w)E_{ij}\otimes E_{kl}\Bigg],
    \end{gathered}
\end{equation}
where
\begin{equation}
    \begin{gathered}
        F_{ij}^a(z)=E_1\Big(z+(-1)^{i-1}\mathbf{q}_{a-1}+(-1)^j\mathbf{q}_a\Big)-
        E_1\Big((-1)^{i-1}\mathbf{q}_{a-1}+(-1)^j\mathbf{q}_a\Big)\,,
    \end{gathered}
\end{equation}
where the indices $i$ for ${\mathbf d}_i(\mathbf{p}_a)$ are modulo $2$ (that is ${\mathbf d}_0={\mathbf d}_2$).

Define the following matrices:
\begin{equation}
    \widetilde{\mathbf{L}}^a(z)=\sum_{i,j=1}^2(-1)^j\mathbf{L}_{ij}^a(z)F_{ij}^a(z)E_{ij}\,,
\end{equation}
\begin{equation}
    \breve{\mathbf{L}}^a(z)=\sum_{i,j=1}^2(-1)^i\mathbf{L}_{ij}^a(z)F_{ij}^a(z)E_{ij}\,,
\end{equation}
and the diagonal matrix
\begin{equation}
    D(p)=\text{diag}(-d(p),d(-p))=\sum_{k=1}^2(-1)^k\mathbf{d}_{k-1}(p)E_{kk}\,.
\end{equation}
\begin{predl}
Then the quadratic $r$-matrix structure for the Lax matrix of the elliptic Toda chain (\ref{w677})
takes the form:
\beq\label{w678}
  \begin{array}{c}
  \displaystyle{
        c\left\{\mathbf{L}^a_1(z),\mathbf{L}_2^b(w)\right\} =
        \delta^{ab}\Big(\mathbf{L}_2^a(w)\mathbf{L}_1^a(z)
\tilde{s}^a_{12}(z)-\mathbf{L}_1^a(z)\mathbf{L}_2^a(w)\tilde{s}^a_{21}(w)\Big)+
        }
        \\ \ \\
    \displaystyle{
        -\delta^{a-1,b}\mathbf{L}^{a-1}_2(w)\mathbf{L}_1^a(z)\tilde{u}^a_{12}(z)+
        \delta^{a,b-1}\mathbf{L}_1^a(z)\mathbf{L}^{a+1}_2(w)\tilde{u}^{a+1}_{21}(w)\,,
    }
    \end{array}
\eq
where $\tilde{s}^a_{12}(z)$, $\tilde{s}^a_{21}(w)$, $\tilde{u}^a_{12}(z)$
and $\tilde{u}^{a+1}_{21}(w)$ defined through
\begin{equation}
    \mathbf{L}_1^a(z)\tilde{s}_{12}^a(z)=
    \widetilde{\mathbf{L}}^a(z)\otimes D(\mathbf{p}_a) = \sum_{i,j,k=1}^2(-1)^{j+k}\mathbf{L}_{ij}^a(z)
    F_{ij}^a(z)\mathbf{d}_{k-1}(\mathbf{p}_{a})E_{ij}\otimes E_{kk}\,,
\end{equation}
\begin{equation}
    \mathbf{L}_2^a(w)\tilde{s}_{21}^a(w)=
    D(\mathbf{p}_a)\otimes\widetilde{\mathbf{L}}^a(w) = \sum_{i,j,k=1}^2(-1)^{j+k}\mathbf{L}_{ij}^a(w)
    F_{ij}^a(w)\mathbf{d}_{k-1}(\mathbf{p}_{a})E_{kk}\otimes E_{ij}\,,
\end{equation}
\begin{equation}
    \mathbf{L}_1^a(z)\tilde{u}_{12}^a(z)=
    \breve{\mathbf{L}}^a(z)\otimes D(\mathbf{p}_{a-1}) = \sum_{i,j,k=1}^2(-1)^{i+k}\mathbf{L}_{ij}^a(z)
    F_{ij}^a(z)\mathbf{d}_{k-1}(\mathbf{p}_{a-1})E_{ij}\otimes E_{kk}\,,
\end{equation}
\begin{equation}
    \mathbf{L}_2^{a+1}(w)\tilde{u}_{21}^{a+1}(w)=
     D(\mathbf{p}_a)\otimes\breve{\mathbf{L}}^{a+1}(w) =
      \sum_{i,j,k=1}^2(-1)^{i+k}\mathbf{L}_{ij}^{a+1}(w)F_{ij}^{a+1}(w)
    \mathbf{d}_{k-1}(\mathbf{p}_{a})E_{kk}\otimes E_{ij}\,.
\end{equation}
\end{predl}

\subsection{Relation to XYZ chain}
Here we use the standard description of the XYZ chain (\ref{w425}), (\ref{w427}) since in contrast to
(\ref{w413})-(\ref{w417}) expressions (\ref{w427}) and (\ref{w429})-(\ref{w432}) have no singularities at $\eta=0$.
Thus, we deal with the Lax matrix
 \beq\label{w4253}
 \begin{array}{c}
  \displaystyle{
{\bf L}(z,{\bf S}^a)=\sigma_0{\bf S}^a_0+\sum\limits_{k=1}^3\sigma_k\vf_k(z){\bf S}^a_k\,,
 }
 \end{array}
 \eq
Plugging $\eta=0$ into (\ref{w429})-(\ref{w432}) we get
\beq\label{w4293}
  \begin{array}{c}
  \displaystyle{
{\bf S}_0^a=\frac12\Big(e^{\bfp_a/2c}+
e^{-\bfp_a/2c}\Big)\,,
}
 \end{array}
 \eq
\beq\label{w4303}
  \begin{array}{c}
  \displaystyle{
{\bf S}_1^a=-\frac12\frac{\theta_4(0)}{\vth'(0)}
\Big(\frac{\theta_4(2\bfq_a)}{\vth(2\bfq_a)}\,e^{\bfp_a/2c}-
\frac{\theta_4(2\bfq_a)}{\vth(2\bfq_a)}\,e^{-\bfp_a/2c}\Big)\,,
}
 \end{array}
 \eq
\beq\label{w4313}
  \begin{array}{c}
  \displaystyle{
{\bf S}_2^a=-\frac{\imath}{2}\frac{\theta_3(0)}{\vth'(0)}
\Big(\frac{\theta_3(2\bfq_a)}{\vth(2\bfq_a)}\,e^{\bfp_a/2c}-
\frac{\theta_3(2\bfq_a)}{\vth(2\bfq_a)}\,e^{-\bfp_a/2c}\Big)\,,
}
 \end{array}
 \eq
\beq\label{w4323}
  \begin{array}{c}
  \displaystyle{
{\bf S}_3^a=-\frac12\frac{\theta_2(0)}{\vth'(0)}
\Big(\frac{\theta_2(2\bfq_a)}{\vth(2\bfq_a)}\,e^{\bfp_a/2c}-
\frac{\theta_2(2\bfq_a)}{\vth(2\bfq_a)}\,e^{-\bfp_a/2c}\Big)\,.
}
 \end{array}
 \eq
The Lax matrices satisfy the classical quadratic exchange relation (\ref{w361}) with the $r$-matrix (\ref{w420})
and provides the same Sklyanin algebra
\beq\label{w4333}
  \begin{array}{c}
  \displaystyle{
c\{{\bf S}^a_i,{\bf S}^a_j\}=\imath\varepsilon_{ijk} {\bf S}^a_0 {\bf S}^a_k\,,
}
\\ \ \\
  \displaystyle{
c\{{\bf S}^a_0,{\bf S}^a_i\}=
\imath\varepsilon_{ijk} {\bf S}^a_j {\bf S}^a_k\big(\wp(\om_j)-\wp(\om_k)\big)
}
 \end{array}
 \eq
as in (\ref{w433}) since the structure constants are independent of $\eta$.

It follows from (\ref{w436}) that in the case $\eta=0$
the Casimir functions
\beq\label{w4343}
  \begin{array}{c}
  \displaystyle{
{\bf C}_1^a=({\bf S}^a_1)^2+({\bf S}^a_2)^2+({\bf S}^a_3)^2\,,
\qquad
{\bf C}_2^a=({\bf S}^a_0)^2+\sum\limits_{k=1}^3 ({\bf S}^a_k)^2\wp(\om_k)
}
 \end{array}
 \eq
take values
\beq\label{w4363}
  \begin{array}{c}
  \displaystyle{
{\bf C}_1^a=0\,,
\qquad
{\bf C}_2^a=1\,.
}
 \end{array}
 \eq
Therefore,
\beq\label{w4353}
  \begin{array}{c}
  \displaystyle{
\det {\bf L}(z,{\bf S}^a)={\bf C}_2^a-\wp(z){\bf C}_1^a=1\,.
}
 \end{array}
 \eq
Then for the monodromy matrix
\beq\label{w520}
  \begin{array}{c}
  \displaystyle{
{\bf T}(z)={\bf L}(z,{\bf S}^1){\bf L}(z,{\bf S}^2)...{\bf L}(z,{\bf S}^n)
}
 \end{array}
 \eq
we obviously have
\beq\label{w521}
  \begin{array}{c}
  \displaystyle{
\det{\bf T}(z)=1\,.
}
 \end{array}
 \eq
However, the XYZ chain related to (gauge equivalent to) the elliptic Toda chain is described by the
modified Lax matrices
\beq\label{w522}
  \begin{array}{c}
  \displaystyle{
{\bf L'}(z,{\bf S}^a)=-\frac{\vth'(0)}{h_{a-1,a}}\,{\bf L}(z,{\bf S}^a)\,,\qquad h_{a-1,a}=b_1^a+b_2^a
}
 \end{array}
 \eq
with $b_1^a$, $b_2^a$ from (\ref{w4761}).
Thus, for the monodromy matrix
%
\beq\label{w523}
  \begin{array}{c}
  \displaystyle{
{\bf T'}(z)={\bf L'}(z,{\bf S}^1){\bf L'}(z,{\bf S}^2)...{\bf L'}(z,{\bf S}^n)
=\frac{(-\vth'(0))^n}{H_0}\,{\bf T}(z)=\frac{(-1)^n}{2^n}\,e^{-H^{\hbox{\tiny{eToda}}}}{\bf T}(z)
}
 \end{array}
 \eq
(with $H_0$ from (\ref{w4765}))
one gets $\det{\bf T'}(z)=\vth'(0)^{2n}/H_0^2=4^{-n}\exp(-2H^{\hbox{\tiny{eToda}}})$ or
\beq\label{w524}
  \begin{array}{c}
  \displaystyle{
\log\det{\bf T'}(z)=-2H^{\hbox{\tiny{eToda}}}-n\log 4\,,
}
 \end{array}
 \eq
which is independent of $z$ and coincides with (\ref{w515}).



%
%

%
%
%

\section{Appendix: elliptic functions}
\def\theequation{A.\arabic{equation}}
\setcounter{equation}{0}

We mainly deal with the elliptic Kronecker function
 \beq\label{a01}
  \begin{array}{l}
  \displaystyle{
 \phi(z,u)=\frac{\vth'(0)\vth(z+u)}{\vth(z)\vth(u)}\,,
 }
 \end{array}
 \eq
 where $\vth(z)$ is the first Jacobi theta-function. In Riemann's notation
 it is as follows.
 Define the theta-functions with characteristics $a,b$:
\beq\label{a02}
 \begin{array}{c}
  \displaystyle{
\theta{\left[\begin{array}{c}
a\\
b
\end{array}
\right]}(z|\, \tau ) =\sum_{j\in \mZ}
\exp\left(2\pi\imath(j+a)^2\frac\tau2+2\pi\imath(j+a)(z+b)\right)\,,\quad {\rm Im}(\tau)>0\,,
}
 \end{array}
 \eq
where $a\,,b\in\frac{1}{N}\,\mZ$.
 In particular, the odd theta function $\vth(z)$ ($\theta_1(z)$ in the Jacobi notation) is
 \beq\label{a03}
 \begin{array}{c}
  \displaystyle{
\vth(z)=\vth(z,\tau)\equiv-\theta{\left[\begin{array}{c}
1/2\\
1/2
\end{array}
\right]}(z|\, \tau )\,.
 }
 \end{array}
 \eq
In the $N=2$ case we also use the Jacobi theta functions:
\beq\label{a031}
\begin{array}{c}
\displaystyle{
\vth(u,\tau)=\theta_1(u|\tau )=-i\sum_{k\in \mZ}
(-1)^k q^{(k+\frac{1}{2})^2}e^{\pi i (2k+1)u},
}
\\ \\
\displaystyle{
\theta_2(u|\tau )\equiv \theta{\left[\begin{array}{c}
1/2\\
0
\end{array}
\right]}(u|\, \tau )
=\sum_{k\in \mZ}
q^{(k+\frac{1}{2})^2}e^{\pi i (2k+1)u},
}
\\ \\
\displaystyle{
\theta_3(u|\tau )
\equiv \theta{\left[\begin{array}{c}
0\\
0
\end{array}
\right]}(u|\, \tau)
=\sum_{k\in \mZ}
q^{k^2}e^{2\pi i ku},
\qquad
\theta_4(u|\tau )
\equiv \theta{\left[\begin{array}{c}
0\\
1/2
\end{array}
\right]}(u|\, \tau)
=\sum_{k\in \mZ}
(-1)^kq^{k^2}e^{2\pi i ku},}
\end{array}
\eq
where $q=e^{\pi i \tau}$.

 The Kronecker function has a single simple pole in variable $z$ at $z=0$:
 \beq\label{a04}
  \begin{array}{l}
  \displaystyle{
\res\limits_{z=0}\phi(z,u)=1
 }
 \end{array}
 \eq
 The following quasi-periodicity properties hold:
 \beq\label{a05}
  \begin{array}{l}
  \displaystyle{
 \phi(z+1,u)= \phi(z,u)\,,\qquad  \phi(z+\tau,u)= \exp (-2\pi\imath u)\phi(z,u)\,.
 }
 \end{array}
 \eq
 The expansion near $z=0$ has the form
 \beq\label{a06}
  \begin{array}{l}
  \displaystyle{
 \phi(z,u)=\frac{1}{z}+E_1(u)+z\,\frac{E_1^2(u)-\wp(u)}{2}+O(z^2),
 }
 \end{array}
 \eq
 where
 \beq\label{a07}
  \begin{array}{l}
  \displaystyle{
 E_1(u)=\frac{\vth'(u)}{\vth(u)}=-E_1(-u)
 }
 \end{array}
 \eq
is the first Eisenstein function.
The relation to the Weierstrass functions is as follows:
\beq\label{a051}
\begin{array}{c}
 \displaystyle{
    E_1(z)=\frac{\vth'(z)}{\vth(z)}=\zeta(z)+\frac{z}{3}\frac{\vth'''(0)}{\vth'(0)}\,,
    \quad
    E_2(z) = - \partial_z E_1(z) = \wp(z) - \frac{\vartheta'''(0) }{3\vartheta'(0)}\,,
}
\end{array}
\eq
that is
\beq\label{a052}
\begin{array}{c}
 \displaystyle{
    \wp(z)=- \partial^2_z\log\vth(z) + \frac{\vartheta'''(0)}{3\vartheta'(0)}\,.
}\end{array}
\eq
The functions $E_1(z)$ and $\wp(z)$ are related by the following identity:
\beq\label{a053}
\begin{array}{c}
 \displaystyle{
E_1(z+w)-E_1(z)-E_1(w)=\frac12\,\frac{\wp'(z)-\wp'(w)}{\wp(z)-\wp(w)}\,.
}
\end{array}
\eq
that is
It follows from the definition (\ref{a01}) that
 \beq\label{a08}
  \begin{array}{l}
  \displaystyle{
 \p_z\phi(z,u)=(E_1(z+u)-E_1(z))\phi(z,u)\,,
 }
 \\ \ \\
   \displaystyle{
  \p_u\phi(z,u)=(E_1(z+u)-E_1(u))\phi(z,u)\,.
 }
 \end{array}
 \eq
A set of the widely known addition formulae (the genus one Fay identity and its degenerations) is used in this paper:
\beq\label{a10}
  \begin{array}{c}
  \displaystyle{
  \phi(z_1, u_1) \phi(z_2, u_2) = \phi(z_1, u_1 + u_2) \phi(z_2 - z_1, u_2) + \phi(z_2, u_1 + u_2) \phi(z_1 - z_2, u_1)
 }
 \end{array}
 \eq
\beq\label{a09}
  \begin{array}{c}
  \displaystyle{
 \phi(z,u_1)\phi(z,u_2)=\phi(z,u_1+u_2)\Big(E_1(z)+E_1(u_1)+E_1(u_2)-E_1(z+u_1+u_2)\Big)\,,
 }
 \end{array}
 \eq
and
\beq\label{a19}
  \begin{array}{c}
  \displaystyle{
  \phi(z, u) \phi(z, -u) = \wp(z)-\wp(u)\,.
 }
 \end{array}
 \eq

For description of XYZ type models we also use the following functions:
 \beq\label{a211}
 \begin{array}{c}
  \displaystyle{
 \vf_{0}(z,x)=\phi(z,x)\,,
\qquad
 \vf_{1}(z,x+\om_1)=e^{\pi\imath z}\phi(z,x+\om_1)=\frac{\theta_1'(0)\theta_{4}(z+x)}{\theta_{1}(z)\theta_{4}(x)}\,,
 }
 \\ \ \\
  \displaystyle{
 \vf_{2}(z,x+\om_2)=e^{\pi\imath z}\phi(z,x+\om_2)=\frac{\theta_1'(0)\theta_{3}(z+x)}{\theta_{1}(z)\theta_{3}(x)}\,,
 }
 \\ \ \\
  \displaystyle{
  \vf_{3}(z,x+\om_3)=\phi(z,x+\om_3)=\frac{\theta_1'(0)\theta_{2}(z+x)}{\theta_{1}(z)\theta_{2}(x)}
 }
  \end{array}
 \eq
 with
 \beq\label{a091}
 \begin{array}{c}
  \displaystyle{
 \om_0=0\,,\quad
 \om_1=\frac{\tau}{2}\,,\quad
 \om_2=\frac{1+\tau}{2}\,,\quad
 \om_3=\frac{1}{2}
  }
 \end{array}
\eq
according to numeration of the Pauli matrices
 \beq\label{a092}
 \begin{array}{c}
  \displaystyle{
 \sigma_0=\mats{1}{0}{0}{1}\,,\quad
  \sigma_1=\mats{0}{1}{1}{0}\,,\quad
  \sigma_2=\mats{0}{-\imath}{\imath}{0}\,,\quad
   \sigma_3=\mats{1}{0}{0}{-1}\,.
  }
 \end{array}
\eq
 Also,
 \beq\label{a27}
 \begin{array}{c}
  \displaystyle{
  \vf_{1}(z)=\frac{\theta_1'(0)\theta_{4}(z)}{\theta_{1}(z)\theta_{4}(0)}\,,\qquad
  \vf_{2}(z)=\frac{\theta_1'(0)\theta_{3}(z)}{\theta_{1}(z)\theta_{3}(0)}\,,\qquad
   \vf_{3}(z)=\frac{\theta_1'(0)\theta_{2}(z)}{\theta_{1}(z)\theta_{2}(0)}\,.
  }
 \end{array}
\eq
For theta functions with $2\tau$ one may use the following identities:
\beq\label{a581}
\begin{array}{c}
   \displaystyle{
\theta_2(x+y|2\tau)\theta_2(x-y|2\tau)=\frac{1}{2}\Big( \theta_3(x|\tau)\theta_3(y|\tau)-
\theta_4(x|\tau)\theta_4(y|\tau) \Big)\,,
 }
 \\ \ \\
    \displaystyle{
\theta_2(x+y|2\tau)\theta_3(x-y|2\tau)=\frac{1}{2}\Big( \theta_2(x|\tau)\theta_2(y|\tau)-
\theta_1(x|\tau)\theta_1(y|\tau) \Big)\,,
 }
  \\ \ \\
   \displaystyle{
\theta_3(x+y|2\tau)\theta_3(x-y|2\tau)=\frac{1}{2}\Big( \theta_3(x|\tau)\theta_3(y|\tau)+
\theta_4(x|\tau)\theta_4(y|\tau) \Big)\,.
 }
\end{array}
\eq
Summation formulae for theta-functions:
\begin{equation}\label{w6231}
\begin{aligned}
&\theta_{1}(u+x)\theta_{1}(u-x)\theta_{r}(v+y)\theta_{r}(v-y) - \theta_{1}(v+x)\theta_{1}(v-x)\theta_{r}(u+y)\theta_{r}(u-y) \\
&= \theta_{1}(u+v)\theta_{1}(u-v)\theta_{r}(x+y)\theta_{r}(x-y), \quad r=1,2,3,4.
\end{aligned}
\end{equation}
\begin{equation}\label{w6241}
\begin{aligned}
&\theta_{2}(u+x)\theta_{2}(u-x)\theta_{3}(v+y)\theta_{3}(v-y) - \theta_{2}(v+x)\theta_{2}(v-x)\theta_{3}(u+y)\theta_{3}(u-y) \\
&= -\theta_{1}(u+v)\theta_{1}(u-v)\theta_{4}(x+y)\theta_{4}(x-y),
\end{aligned}
\end{equation}
\begin{equation}\label{w6251}
\begin{aligned}
&\theta_{2}(u+x)\theta_{2}(u-x)\theta_{4}(v+y)\theta_{4}(v-y) - \theta_{2}(v+x)\theta_{2}(v-x)\theta_{4}(u+y)\theta_{4}(u-y) \\
&= -\theta_{1}(u+v)\theta_{1}(u-v)\theta_{3}(x+y)\theta_{3}(x-y),
\end{aligned}
\end{equation}
\begin{equation}\label{w6261}
\begin{aligned}
&\theta_{3}(u+x)\theta_{3}(u-x)\theta_{4}(v+y)\theta_{4}(v-y) - \theta_{3}(v+x)\theta_{3}(v-x)\theta_{4}(u+y)\theta_{4}(u-y) \\
&= -\theta_{1}(u+v)\theta_{1}(u-v)\theta_{2}(x+y)\theta_{2}(x-y).
\end{aligned}
\end{equation}
\begin{equation}\label{w6271}
\begin{aligned}
&\theta_{r}(u+x)\theta_{r}(u-x)\theta_{r}(v+y)\theta_{r}(v-y) - \theta_{r}(u+y)\theta_{r}(u-y)\theta_{r}(v+x)\theta_{r}(v-x) \\
&= (-1)^{r-1} \theta_{1}(u+v)\theta_{1}(u-v)\theta_{1}(x+y)\theta_{1}(x-y), \quad r=1,2,3,4.
\end{aligned}
\end{equation}


\paragraph{Acknowledgments.}
We are grateful to A. Zabrodin for useful discussions.

The work of A. Zotov was performed at the Steklov International Mathematical Center and supported by the Ministry of Science and Higher Education of the Russian Federation (agreement no. 075-15-2025-303).





\begin{small}

\end{small}

\end{document}